\documentclass{aastex62}

\usepackage{amsmath}
\usepackage{chemfig}
\shorttitle{Nitrogen ice mixture}
\shortauthors{Ciaravella et al.}
\received{January 2019}
\revised{April 2019}
\accepted{May 2019}

\begin{document}
\title{\LARGE Synthesis of complex organic molecules in soft X-ray irradiated ices}
\correspondingauthor{Y.-J. Chen}
\email{asperchen@phy.ncu.edu.tw}

\author{A. Ciaravella$^1$, A. Jim\'enez-Escobar$^1$, C. Cecchi-Pestellini}
\affil{INAF - Osservatorio Astronomico di Palermo, P.za Parlamento 1, 90134 Palermo, Italy}
\nocollaboration

\author{C.-H., Huang$^2$, N.-E., Sie}
\affiliation{Department of Physics, National Central University, Jhongli City, Taoyuan County 32054, Taiwan}
\nocollaboration

\author{G.M. Mu\~{n}oz Caro}
\affiliation{Centro de Astrobiolog\'ia (INTA-CSIC), Carretera de Ajalvir, km 4, Torrej\'on de Ardoz, 28850 Madrid, Spain}

\author{Y.-J. Chen}
\altaffiliation{Department of Physics, National Central University, Jhongli City, Taoyuan County 32054, Taiwan}

\begin{abstract}
We study the chemical evolution of H$_2$O:CO:NH$_3$ ice mixtures irradiated with soft X-rays, in the range $250-1250$~eV. We identify many nitrogen-bearing molecules such as e.g., OCN$^-$, NH$_4^+$, HNCO, CH$_3$CN, HCONH$_2$, and NH$_2$COCONH$_2$. Several infrared features are compatible with glycine or its isomers.

During the irradiation,  we detected through mass spectroscopy many species desorbing the ice. Such findings support either the infrared identifications and reveal less abundant species with not clear infrared features. Among them, $m/z=57$ has been ascribed to methyl isocyanate (CH$_3$NCO), a molecule of prebiotic relevance, recently detected in protostellar environments. 

During the warm up after the irradiation, several infrared features including 2168~cm$^{-1}$ band of OCN$^-$, 1690 cm$^{-1}$ band of formamide, and the 1590 cm$^{-1}$ band associated to three different species, HCOO$^-$, CH$_3$NH$_2$ and NH$_3^+$CH$_2$COO survive up to room temperature. Interestingly, many high masses have been also detected. Possible candidates are methyl-formate, ($m/z=60$, HCOOCH$_3$), ethanediamide ($m/z=88$, NH$_2$COCONH$_2$), and  N-acetyl-L-aspartic acid ($m/z=175$). This latter species is compatible with the presence of the $m/z=43$, 70 and 80 fragments. 

Photo-desorption of organics is relevant for the detection of such species in the gas-phase of cold environments, where organic synthesis in ice mantles should dominate. We estimate the gas-phase enrichment of some selected species in the light of a protoplanetary disc model around young solar-type stars.
\end{abstract}
\section{Introduction}\label{s1}
Ice mantles covering cold dust particles in dense clouds and circumstellar regions are laboratories for chemistry giving rise to a plethora of chemical species \citep{Her09}. Processing of space ice analogues with ultraviolet photons \citep{Ber95,MC02,Ber02,Mu03,M04,N06,Ch07,dM11,V13,MCD13,M16,Ob16}, and energetic particles \citep{P98,L05,Sic12,I14,MC14,J14,F16} have widely explored the cold solid state synthesis of complex organic molecules. Formation routes, mechanisms responsible for the production of new species, and efficiency of the different energetic processing have contributed to the interpretation  of the  observed interstellar ice features \citep{Bo15}. More recently X-ray processing of ice analogues has been studied given its relevance in circumstellar discs around young stars \citep{C10,A10,C12,J12,Ch13,PB15,C16,Jim16,Jim18}. In solar type stars X-ray emission dominates over far extreme ultraviolet emissions for almost one billion year \citep{R05} penetrating through the disc and reaching inner regions otherwise forbidden to less energetic photons (e.g., \citealt{W12}). 

In this paper we present evidence for the synthesis of N-bearing organic molecules induced by soft X-rays in an H$_2$O:CO:NH$_3$ (1:0.9:0.7) ice mixture. The photo-desorption of a similar mixture is discussed in \citet{Jim18}. 

The photochemistry of ices containing ammonia was first studied in astrophysically relevant ice mixtures by \citet{H79}, who irradiated a CO ice containing traces of water, ammonia and carbon dioxide. This early investigation and others that followed ~\textendash~ changing the initial composition, the abundance ratios, and the energy source ~\textendash~ had as major outcomes the formations of XCN compounds (in particular OCN$^-$), and moderately complex amino containing molecules, which may be precursors for amino acids. The largest infrared detected molecule was $\rm C_6H_{12}N_4$ \citep{Mu03}. Subsequently, two independent studies formed amino acids by zapping dirty water ices with ultraviolet radiation \citep{Ber02,MC02}. The ices contained a fairly high amount of ammonia, methanol and hydrogen cyanide. In addition, the processing of nitrogen bearing mixtures produced species with the peptide moiety, such e.g., isocyanic acid, HNCO, the smallest stable molecule containing all four primary biogenic elements, and formamide, $\rm NH_2CHO$ (e.g., \citealt{Dem98,J11,Ji14,K16}). Among the energy sources available in space, ultraviolet radiation and cosmic rays (e.g., \citealt{Hu00,Pil10}) have been the most exploited in this kind of studies. 

There are certainly differences in the way different energy sources affect the chemistry. Depending on the incident photon energy, different set of orbitals become energetically possible. Soft X-rays are  particularly effective in ionizing core levels. If the core-level of an atom or a molecule is photoionized, the ionic state is highly unstable and will decay to a state of lower energy. In (relatively) light atoms ($Z \la 30$), when an electron is removed from the atom core level (photo-electron), a higher energy level electron may fall in the vacancy. The excess energy involved causes the emission from the atom of another electron known as Auger electron. The kinetic energy of the Auger electron is defined by the difference between the intermediate (core-ionized state) and final states. During the photoexcitation of a core-level, the electron is first promoted to an unoccupied orbital, with resulting core-hole filled by an electron as in the normal Auger case. Then, an Auger electron is released with an energy corresponding to the excess energy of the transition.

The ejected electrons interact with the ice, producing a chain of (valence orbital) ionizations, freeing gradually other electrons of increasingly smaller energy, that, until they can, keep ionizing the ice material. This leads to the fragmentation of the ice molecules with a creation of a large number of ions and radicals, through a number of processes such as e.g., dissociative electron attachment, in which a molecule captures a low-energy electron in an excited resonant state, forming a transient molecular anion, that eventually dissociates (e.g., \citealt{M12}).

We describe the experiment in Section~\ref{exp}. We present and discuss 
the results in Section~\ref{res}: irradiation products, the warm up 
phase followed through infrared and mass spectroscopy, and the 
refractory residue left on the window after the warm up of the sample to
room temperature. The conclusions and the astrophysical implications are
in Section~\ref{disc}.
\section{Experiment}\label{exp}
\begin{figure}
\centering
\includegraphics[width=9cm]{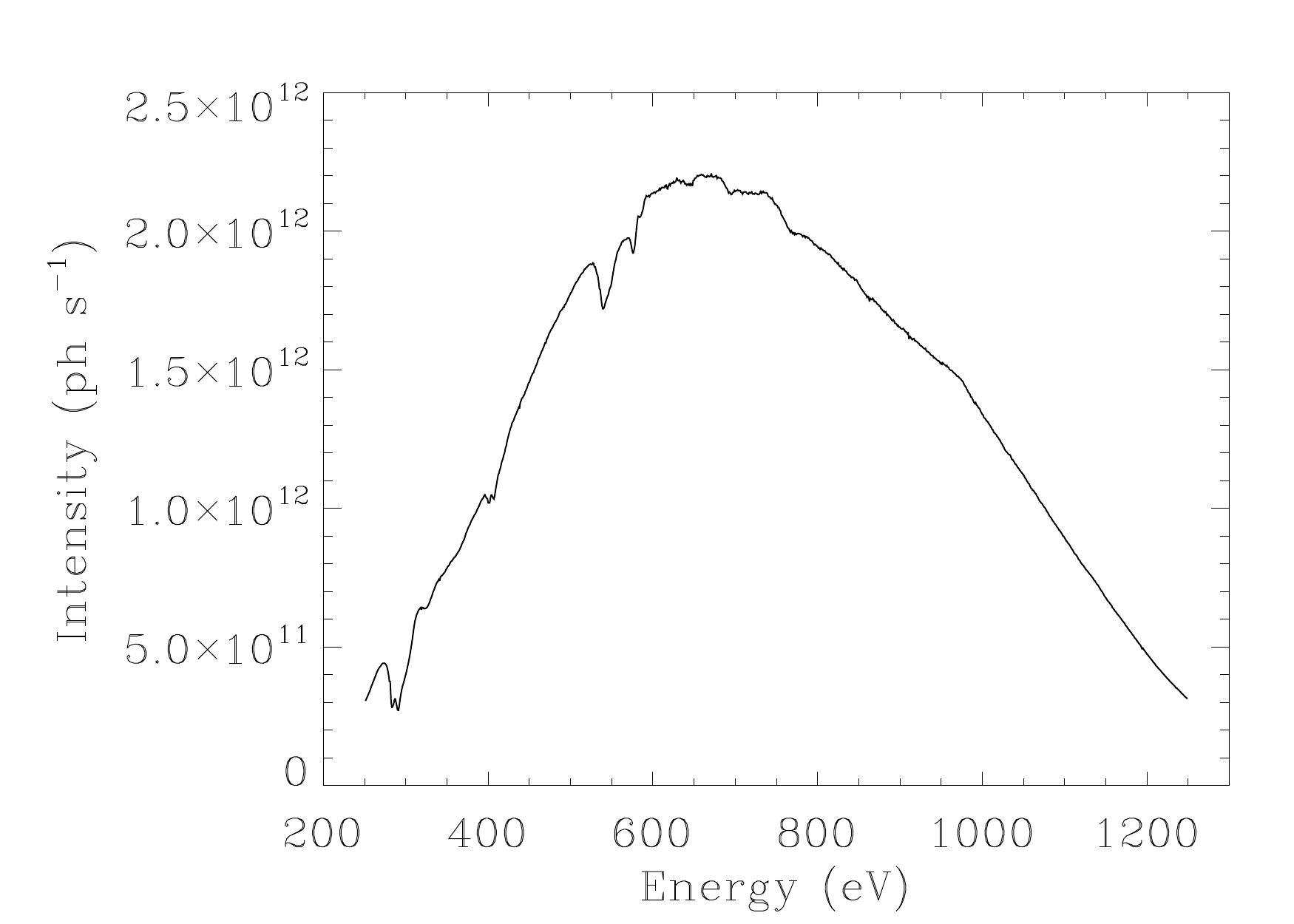} 
\caption{X-ray spectrum of the BL08B beam-line at NSRRC used for the 
irradiation.}
\label{f1}
\end{figure}

The experiments were carried out in the Interstellar Photo-process System (IPS), an ultrahigh vacuum (UHV) chamber of base pressure $1.3 \times 10^{-10}$~mbar. A mid-infrared Fourier Transform Infrared ABB FTLA-2000-104 spectrometer equipped with a mercury-cadmium-telluride infrared detector records infrared spectra in transmission of the ice sample. A Quadrupole Mass Spectrometer (QMS) in the $1-200$~amu range (0.5 amu resolution) were used to monitor the ice and the composition of the gas in the chamber during the whole experiment. The gas line system baked out at $120^\circ$C to eliminate organic and water contamination reaches a minimum pressure of 
$1.3 \times 10^{-7}$ mbar before preparing the gas mixture for the experiments. For a detailed description of the IPS facility see \citet{Ch14}. As X-rays source we used the Spherical Grating Monochromator beamline BL08B at National Synchrotron Radiation Research Center (NSRRC, Taiwan) covering photon energies from 250 to 1250 eV, whose spectrum is in Figure~\ref{f1}. During the experiments, X-ray photon flux is monitored by an in-line nickel mesh (about 90\% optical transmission), calibrated by a traceable photodiode (International Radiation Detectors, Inc.).

A KBr window cooled to 13 K  was used as substrate for a H$_2$O:CO:NH$_3$ (1:0.9:0.7) ice mixture. Such a mixture was irradiated with the soft X-ray spectrum of Figure~\ref{f1} for a total of 120 min. In the mixture we used H$_2$O from Merck, GC-mass grade, freeze-pump-thaw degassing more than 3 times;  CO and NH$_3$ from Matheson, 99.99\% purity. The column densities of the ice components were 8.2 $\times$ 10$^{17}$ molecules cm$^{-2}$ for H$_2$O, derived using the 5018~cm$^{-1}$ band with a band strength $A \rm (H_2O) = 1.2 \times 10^{-18}$~cm~molecule$^{-1}$ \citep{Ger05}, $7.0 \times 10^{17}$ molecules cm$^{-2}$ for CO using the band at 2142 cm$^{-1}$ with $A \rm (CO) = 1.1 \times 10^{-17}$~cm~molecule$^{-1}$ \citep{Jia75}, and $5.5 \times 10^{17}$ molecules cm$^{-2}$ for NH$_3$ using the band at 1112 cm$^{-1}$ with $A \rm (NH_3) = 1.7 \times 10^{-17}$~cm~molecule$^{-1}$ \citep{San93}. The total column density of the mixture is $N ({\rm CO}) + N ({\rm H_2O}) + N({\rm NH_3}) = 2.1 
\times 10^{18}$ molecules cm$^{-2}$.

Before and after irradiation, infrared spectra were collected with a resolution of 1, 2, and 4~cm$^{-1}$. The irradiation was stopped ten times (0.5, 1, 5, 10, 20, 40, 60, 80, 100, 120 min) and infrared spectra were taken with a resolution of 2 cm$^{-1}$. end of the irradiation, the ice was heated up to room temperature at a rate of  2 K minute$^{-1}$. During the warm-up, infrared spectra were acquired every 10 K with a resolution of 4 cm$^{-1}$. The QMS was scanning masses from 1 to 200 amu during the whole experiment. We also ran a blank experiment using a similar ice mixture and no irradiation.

\begin{figure}
\centering
\includegraphics[width=9cm]{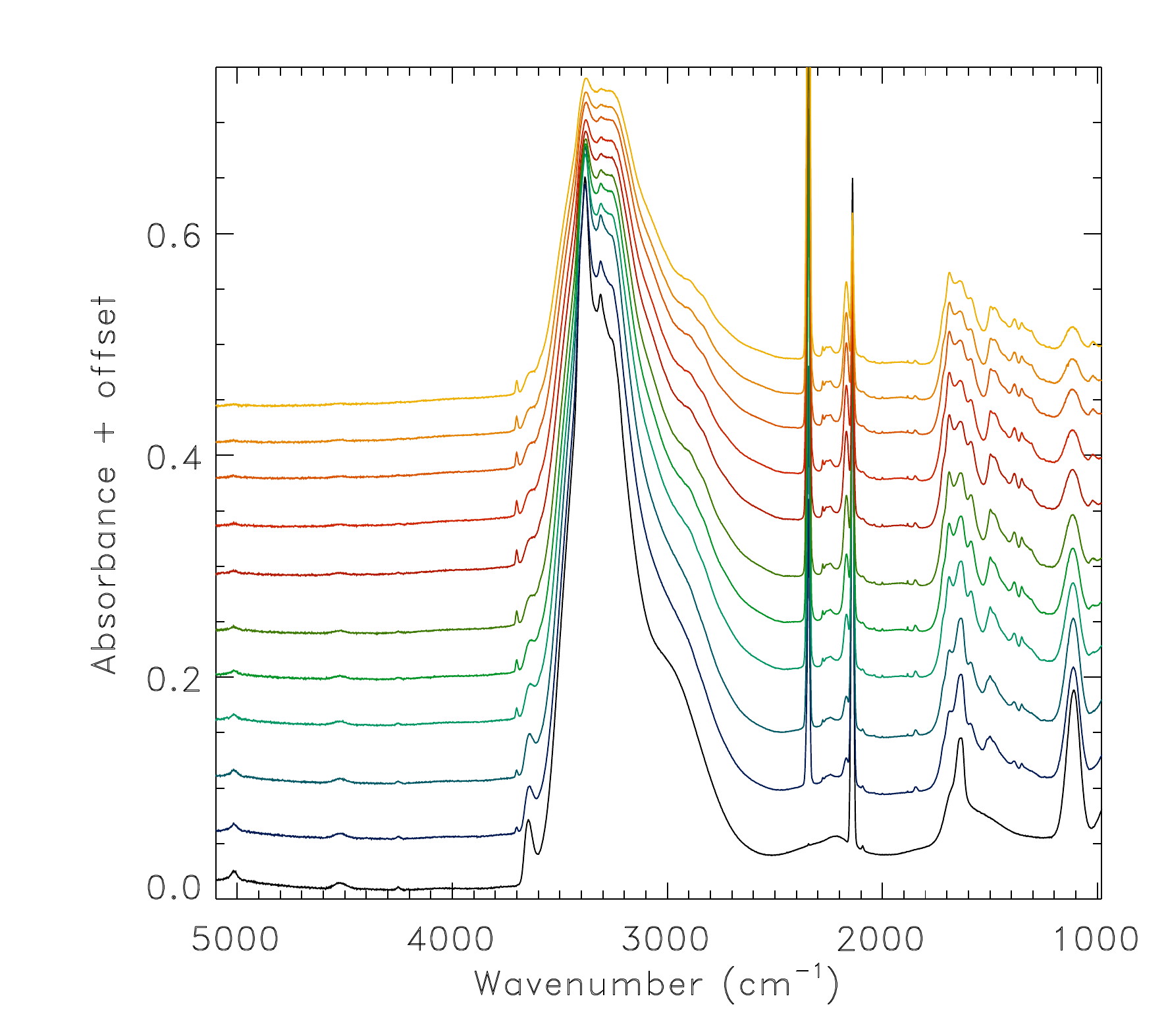} 
\caption{Infrared spectra of the mixture in the range $980 - 3800$~cm$^{-1}$. From the bottom the curves are the spectra of the ice before irradiation followed by those  taken after 0.5, 1, 5, 10, 20, 40, 60, 80, 100, and 120 min irradiation. The curves are shifted for clarity.}
\label{f2}
\end{figure}

\section{Results and Discussion}\label{res}
\subsection{Irradiation and Products}\label{prod}
During the irradiation infrared spectra and mass spectra of the photo-desorbing species have been recorded. It is worthy to mention that both techniques have a degree of degeneracy. Blending of the infrared features of different species prevent in many case a unique identification. Mass spectra of less abundant and/or heavier species are more difficult to detect. We used infrared and mass spectra to identify the products and whenever necessary we based our identification on the most probable formation routes during the irradiation and the warm up.

The primary effect of the X-rays is the core ionization of the atoms in the molecule as
\begin{equation}
\label{eq1}
\rm
H_2O \xrightarrow{\it h\nu} H_2O^{+\star} +  e^-_{ph} \to OH^+ + H^+ + e^-_{ph} + e^-_A 
\end{equation}
\citep{Ma98}, and
\begin{equation}
\label{eq3}
\rm
CO \xrightarrow{\it h\nu} CO^{+\star} + e^-_{ph} \to
\left \{
\begin{tabular}{ll}
CO$^{2+}$ + e$^-_{ph}$ + e$^-_A$ \\
C$^+$ + O$^+$ + e$^-_{ph}$ + e$^-_A$ 
\end{tabular}
\right.
\end{equation}
\citep{Be99}. The star symbol (*) indicates an excited state. In a liquid solution of ammonia and water a nitrogen 1s core-level ionization results in 
\begin{equation}
\label{eq2}
\rm
NH_3 \xrightarrow{\it h\nu} NH_2^{\star} + H^+ + e^-_{ph} 
\end{equation}
\citep{U15}. In the case of N~$1s$ excitation to the $4a1$ molecular orbital, the core-excited molecule de-excites  primarily via resonant Auger decay into
\begin{equation}
\label{eq3}
\rm
NH_3 \xrightarrow{\it h\nu}  NH_3^{\star}  \to
\left \{
\begin{tabular}{ll}
NH$_2^{+}$ + H$^+$ + 2 e$^-_A$ \\
NH$^{+}$ + H$^+$ + H + 2 e$^-_A$
\end{tabular}
\right.
\end{equation}
The ejected primary photo-electrons, $e^-_{ph}$ and Auger electrons, $e^-_A$ will further interact with the ice, giving rise to a cascade of secondary electrons that will drive the chemistry.

\begin{figure*}
\centering
\includegraphics[width=18cm,height=10cm]{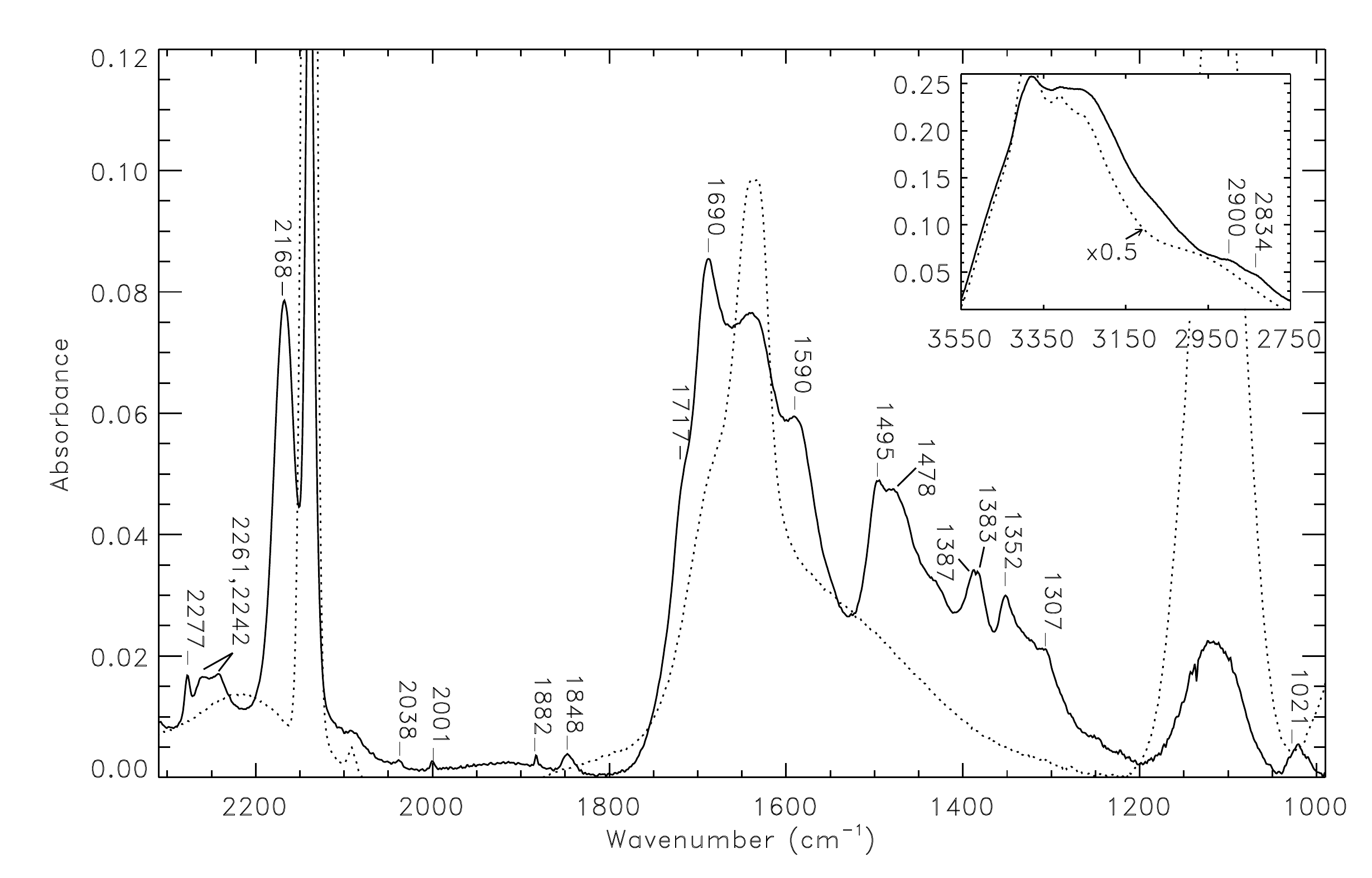} 
\caption{Infrared spectra of the mixture in the range $980-2310$~cm$^{-1}$ before (dotted line) and after irradiation (solid line).  The wavenumber of the main products of the irradiation are marked. The inset panel shows the spectra in the range $2750 - 3550$~cm$^{-1}$. In this case the sample spectrum (dotted line) has been divided by 2.}
\label{f3}
\end{figure*}

\begin{figure*}[ht!]
\centering
\hspace{-2cm}
\begin{tabular}{cc}
\includegraphics[width=9cm]{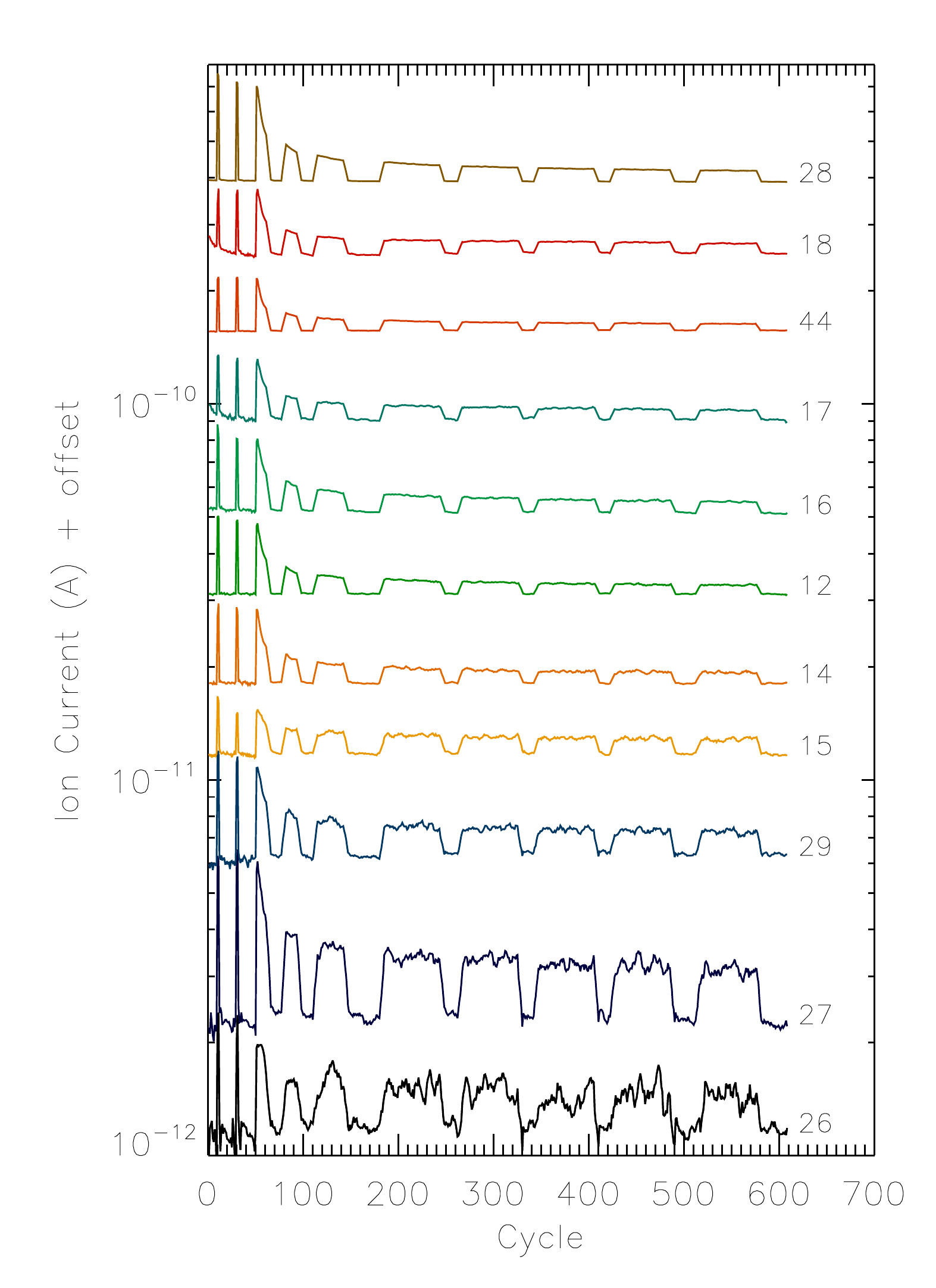} 
\includegraphics[width=9cm]{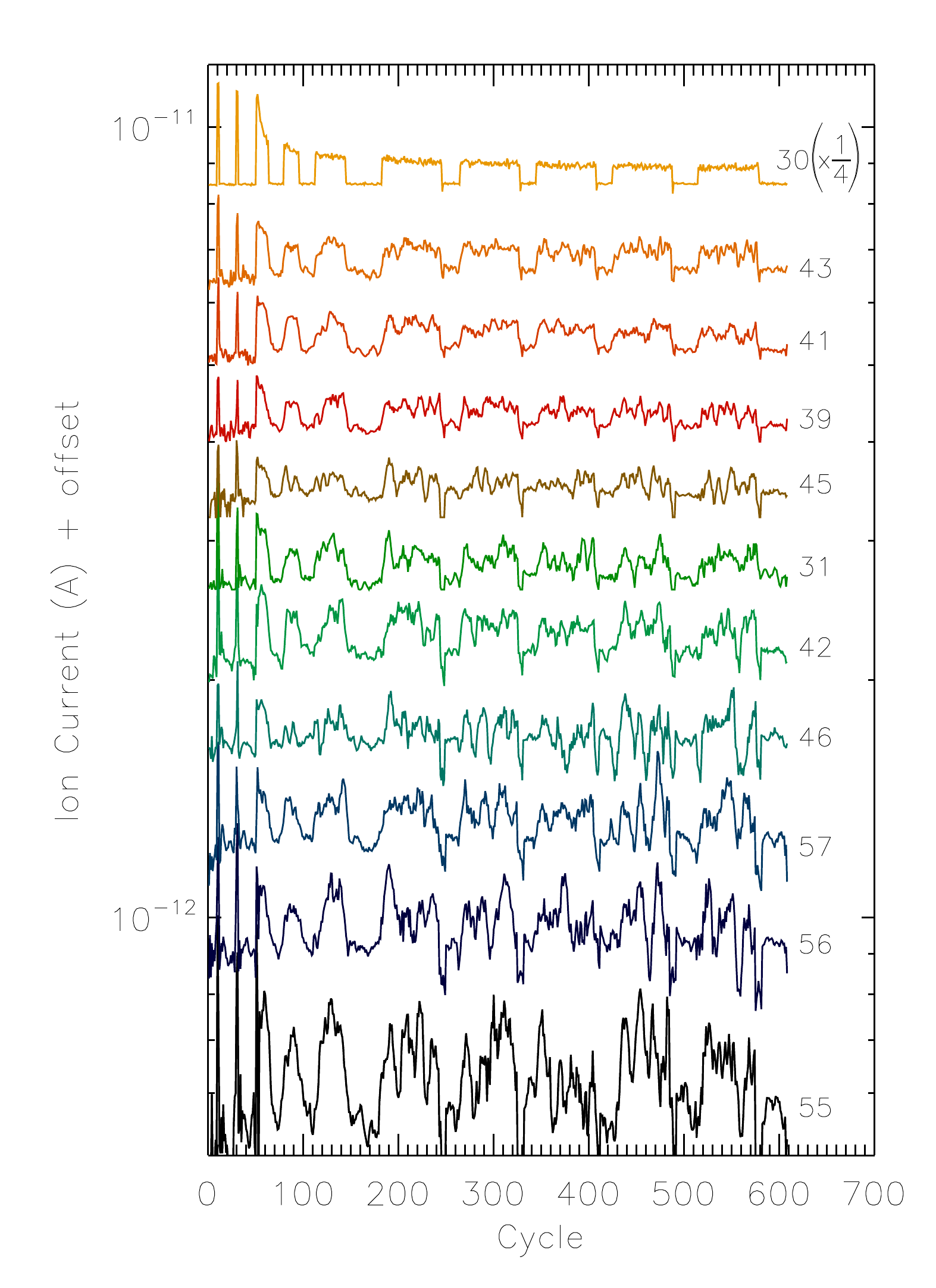} 
\end{tabular}
\caption{Ion current vs cycle as detected by the QMS during the irradiation. The curves are all shifted for clarity. Left panel:  $m/z= 12$, 14,15, 16, 17, 18, 26, 27, 28, 29, and 44. Right panel: $m/z= 30$, 31, 39, 41, 42, 43, 45, 46, 55, 56 and 57. The ion current for $m/z=30$ has been divided by 4.}
\label{f4}
\end{figure*}

In Figure~\ref{f2} are shown the infrared spectra of the mixture sample and those taken at different steps during the irradiation. From Figure~\ref{f2}, the range between $2400 - 1000$~cm$^{-1}$ is the richest in new features. A close-up view of this portion along with the region between $2750-3550$~cm$^{-1}$ is reported in Figure~\ref{f3}, in which we 
marked the most abundant species (see Table~\ref{t1}). The spectral region from 1800 to 1000~cm$^{-1}$ is dominated by NH$_4^+$, -NH$_3^+$, and -COO$^-$ stretchings and C-H~bending, found in a large variety of organic molecules, from simplest compounds to more
complex molecules such as amino acids.
In Figure~\ref{f4} are shown the mass spectra of the main fragments desorbing during the irradiation. Along with the masses of the parent molecules (CO $m/z=28$, H$_2$O $m/z=18$ and NH$_3$ $m/z=17$) and their main fragments (having 10 desorption peaks as many as the number of the irradiation steps), we also detected many other molecules showing the same pattern. Although more noisy, less abundant species/fragments show similar pattern.

The majority of the infrared features appear after the first 30~s irradiation. CO$_2$ (2342 cm$^{-1}$, $m/z=44$) is the most abundant product. HNCO (2168 cm$^{-1}$) is a second generation product formed by direct reaction of CO with NH$_2$ produced after the direct interaction of NH$_3$ with X-ray (reaction~\ref{eq2}, see also \citealt{Hu00})
\begin{equation}\label{eq4}
\rm
NH_2 + CO \rightarrow HNCO + H
\end{equation}

OCN$^-$ (2168 cm$^{-1}$) arises from the reaction between NH$_3$ and HNCO
\begin{equation}\label{eq5}
\rm NH_3 + HNCO \xrightarrow{NH_3} OCN^- + NH_4^+
\end{equation}

The cyanate ion has been detected in many experiments of ice irradiation and radiolysis \citep{Ger04,Br05,Ch07,Pil10}, and also observed in many astrophysical sources \citep{Ob11,Aik12}. The features at 2261 and 2242~cm$^{-1}$ are associated to HNCO \citep{Rau04}. The detection of mass $m/z=43$ during the irradiation and the warm up support the presence of this species. In this region of the spectrum only a band at 2233 cm$^{-1}$  has been detected in 46 MeV Ni ions irradiation of a similar mixture by \citet{Pil10}, and has been assigned to N$_2$O. 

At 2260 cm$^{-1}$ there is the $\nu_2$ CN-stretch feature of acetonitrile (CH$_3$CN) \citep{Men16}. Mass spectra of $m/z=41$ have been detected during irradiation and warm up. Such mass its is also associated to an isomer of acetonitrile CH$_2$CNH compatible with the infrared feature at 2038 cm$^{-1}$ as listed in Table~\ref{t1}. Both species have been detected in space \citep{Mcg18}.

The band at 1690 cm$^{-1}$ and the shoulder at 2900 cm$^{-1}$ can be associated to formamide, HCONH$_2$ (see Figure~\ref{f3}). The other bands of formamide at 3368 and 3181 \citep {Br06} are blended under the broad feature of $\rm H_2O + NH_3$. Photo-desorption of  mass $m/z=45$, although is mostly associated to $\rm ^{13}CO_2$, might have been contributed by formamide. HCONH$_2$ can be formed by reaction of NH$_2$ with HCO
\begin{equation}\label{eq6}
\rm
NH_2 + HCO \rightarrow HCONH_2
\end{equation}
HCO is easily formed by the interaction of CO with H formed during irradiation. For further details see the chemical reaction scheme in \citet{Hu00}. The feature at 1495 cm$^{-1}$, is assigned to NH$_4^+$ and/or NH$_2$. Their formation takes place via NH$_3$ photolysis and subsequent addition of H$^+$ produced by either H$_2$O or NH$_3$ reactions in equations~\ref{eq1} and \ref{eq2} \citep{Pil10,Hu00}, or other channels provided by the secondary electron cascade
\begin{equation} \label{eq7}
\rm
NH_3 + H^+ \rightarrow NH_4^+
\end{equation}    
Another possible proton donor could the hydronium ion H$_3$O$^+$, how it occurs in liquid solution.

The absorption at 1590 cm$^{-1}$ is widely contributed by the vibrational modes of the asymmetric stretching of HCOO$^-$ and the -NH$_2$ scissor of species such as CH$_3$NH$_2$  ($m/z = 31$) \citep{Hol05}, an intermediate in the amino acids formation \citep{Wo02} or even an amino acid such as glycine \citep{Mat18}. The features at 1495 and 1590 together with the feature at 2900  cm$^{-1}$ are also associated to the zwitterionic , NH$_3^+$CH$_2$COO$^-$ \citep{Hu00}, or the neutral form of glycine, NH$_2$CH$_2$COOH \citep{Go03, Ma11}. Formation of glycine is also supported by QMS detection of $m/z=75$ desorbing in the range $200-240$~K (see Figure~\ref{f9}) during the warm-up. Glycine formation in ice mixtures have been inferred in ion irradiation \citep{Pil10}. However, as remarked by \citet{Ob16} in ultraviolet-irradiated ice analogs, glycine is not a major component of $m/z=75$, with many other isomers as methylcarbamic acid, CH$_3$NHCOOH, and glycolamide, NH$_2$COCH$_2$OH, that dominate.

\begin{figure}[!h]
\centering
\includegraphics[width=9cm]{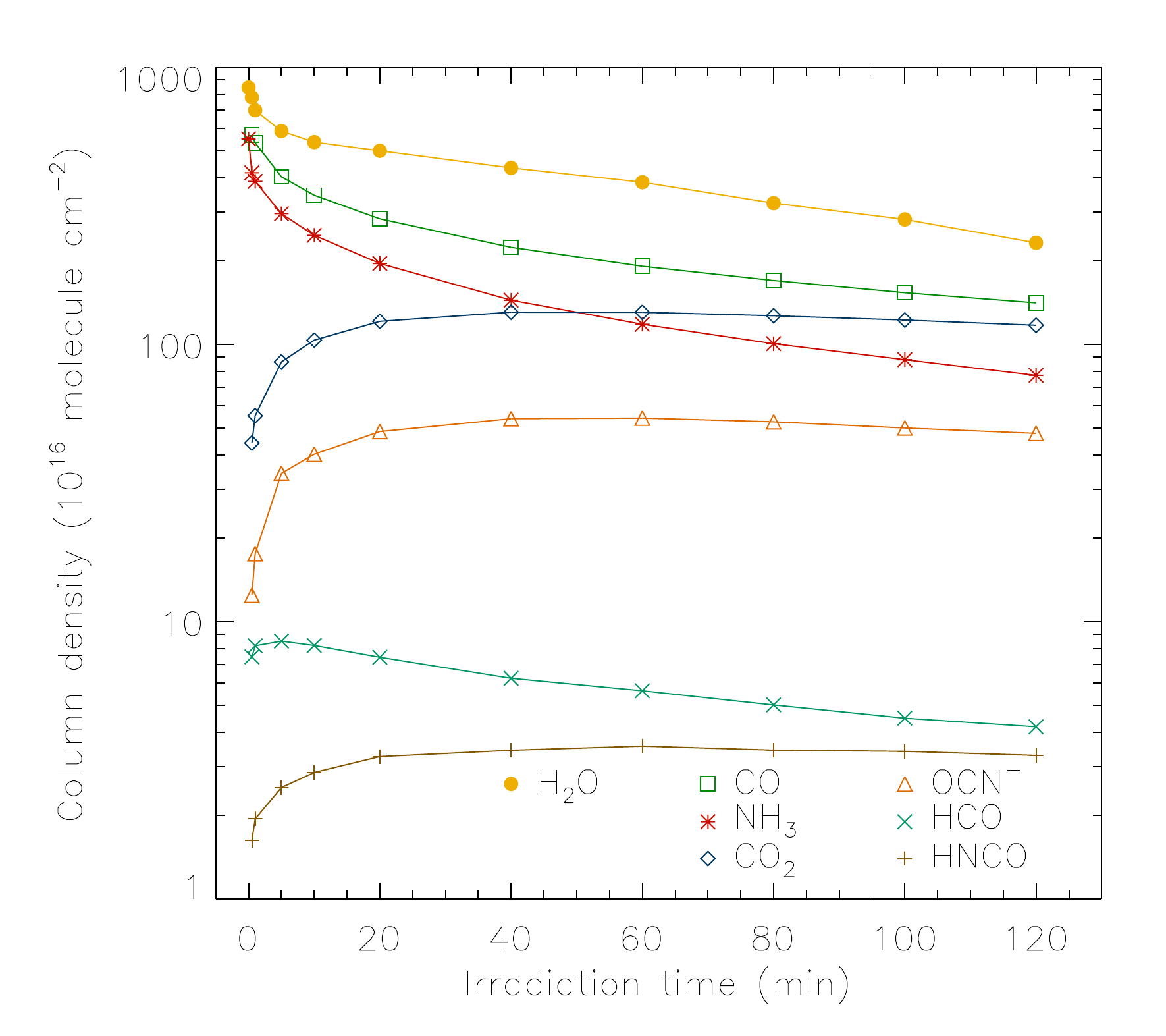} 
\caption{Column density as function of the irradiation time for the parent species H$_2$O (yellow dots), CO (green squares) and NH$_3$ (red asterisks), and the products CO$_2$ (blue diamonds), OCN$^-$ (orange triangles), HCO (light green crosses), and HNCO (light brown pluses).}
\label{f5}
\end{figure}

Table~\ref{t1} summarizes  all the infrared features identified in the spectra, their assignments, and for some species the band strengths. In Figure~\ref{f5} we report the  column densities of H$_2$O, CO and NH$_3$ and four products CO$_2$ 2342  cm$^{-1}$, OCN$^-$ 2168~cm$^{-1}$, HCO 1848 cm$^{-1}$ and HNCO 2242, 2261 cm$^{-1}$. Two Gaussian fitting is used to compute the integrated absorbances of CO and OCN$^-$. HNCO is the sum of the 2241 and 2261 cm$^{-1}$ bands \citep{Rau04}. Three Gaussian fitting needs to separate the bands of HNCO from $\rm ^{13}CO_2$ at 2277 cm$^{-1}$.
At the end of the irradiation $\sim 73$\% of the H$_2$O, $\sim 75$\% of the CO and $\sim 85$\% of NH$_3$ have been destroyed. CO$_2$, OCN$^-$ and HNCO increase for the first 60 min irradiation. OCN$^-$ species is always less abundant than CO$_2$. HCO increases  during the first 5 min and then decreases. This species was not detected in heavy ions irradiation \citep{Pil10}.
\begin{deluxetable}{lll}[!h]
\tabletypesize{\scriptsize} 
\tablecaption{Identification of the Products of X-Ray Irradiation \label{t1}}
\tablewidth{0pt}
\tablehead{Band & Assignment & Band Strength \\
(cm$^{-1}$) &  & (cm molec$^{-1}$)}
\startdata
 3700 &  CO$_2$  &  \nodata\\
 3380 & NH$_3$ &  \nodata\\
 2900 &  HCONH$_2^{a}$ & \nodata \\
 2834 & ?? weak & \nodata\\
 2342 & CO$_2$                              & 7.6$\times$10$^{-17b}$  \\
 2277 & $^{13}$CO$_2$                  &  7.8$\times$10$^{-17c}$ \\ 
 2261 & HNCO, CH$_3$CN$^d$  & 7.2 $\times$10$^{-17e}$\\
 2242 &  HNCO, N$_2$O (?)               &  \nodata\\ 
 2168 & OCN$^-$                            &  1.3$\times$10$^{-16c}$\\
 2139 & CO                                      &  1.1$\times$10$^{-17}$\\ 
 2038 &  C$_3^f$, CH$_2$CNH$^g$,HNC$^h$   & \nodata  \\
 2001 &   C$_2$O$^i$, CNNC$^j$                      &  \nodata  \\ 
 1882 & CO$_3^i$, NCO$^j$, NO$_X$          & 4.5$\times$10$^{-18j}$\\
 1848 & HCO$^m$, HOCO, N$_2$O$_3^l$& 9.6$\times$10$^{-18m}$\\
 1841& HCO, HOCO & \nodata\\
 1717 &  HCOOCH$_3^n$,  H$_2$CO & 4.9$\times$10$^{-17n}$    \\
          & HCOOH$^u$, H$_2$CO$_3$,NH$_2$CH$_2$COOH$^{-o}$    &  \nodata \\
 1690 &  HCONH$_2^{c,a}$                    & \nodata  \\ 
 1638 &  HCOOH$^u$, NH$_3$, N$_2$O$_3^p$       &  \nodata \\
 1590 & HCOO$^-$, CH$_3$NH$_2^q$, NH$_3^+$CH$_2$COO$^{-q}$ & \nodata \\
 1495 & NH$_2^r$, NH$_4^{+s}$, NH$_3^+$, NH$_3^+$CH$_2$COO$^{-q}$& \\
 1478 &   NH$_4^+$,  NO$_3^u$(?) & \\ 
 1387 & -COO$^-$,HCONH$_2^{a}$                         & \nodata \\
 1383 & ?? & \\ 
 1352 & CH$_3$CHO, HCOO$^-$, NH$_3^+$CH$_2$COO$^-$ &  \nodata\\
 1307 & CH$_4$, N$_2$O$_3^s$, N$_2$O$_4^{p,s}$,HCONH$_2$?   & \\
 1250 & NH$_2$CH$_2$COOH$^{o}$ & \\
 1112 & NH$_3^t$  ....                            & 1.7$\times$10$^{-17t}$\\
 1021& CH$_3$OH$^b$, O$_3$?          & 1.8$\times$10$^{-17b}$ \\
\enddata
\tablecomments{
$^a$\citet{Br06},
$^b$\citet{Ger96},
$^c$\citet{Bro04},
$^d$\citet{Men16},
$^e$ \citet{The11}, 
$^f$\citet{Hin88},
$^g$\citet{Men16},
$^h$\citet{Men18},
$^i$\citet{Jim16},
$^j$\citet{Sic12},
$^k$\citet{Che15},  
$^l$\citet{Jam05},
$^m$ HCO; \citet{Mil71},  
$^n$\citet{Mod10},
$^o$\citet{Go03},
$^p$\citet{Var71},  
$^q$\citet{Hol05},  
$^r$\citet{Zhe08},  
$^s$\citet{Pil10},
$^t$\citet{San93}
$^u$\citet{Bis07}}
\end{deluxetable} 

During the irradiation, thanks to its high sensitivity, the QMS detected masses of low abundant species that have no clear infrared features. As an example $m/z = 55$, 56 and 57 cannot be related to species listed in Table~\ref{t1}. A promising candidate for $m/z=57$ is methane isocyanate (CH$_3$NCO). The most intense feature of this species at 2278 cm$^{-1}$ \citep{Ma17} overlaps  with the  $^{13}$CO$_2$. However the detections of $m/z = 57$ along with the two fragments CH$_2$NCO ($m/z=56$) and CHNCO ($m/z=55$) support the presence of such species. This molecule has been recently detected in a low-mass solar-type protostellar binary \citep{Mar17,Lig17}.

Following the analysis by \citet{Jim18} we identified the species associated to the masses detected during the irradiation, see Table~\ref{t2}. In the same table we also list the masses detected during the subsequent warm-up phase along with their possible candidate species.

\begin{deluxetable}{lccl}
\tabletypesize{\scriptsize} 
\tablecaption{QMS detected fragments \label{t2}}
\tablewidth{0pt}
\tablehead{$m/z$ & Irrad. & Warmup &Fragments}
\startdata
12 & y & y & C\\
13 & - & y &  $^{13}$C \\
14 & y & y &  N, CH$_2$ \\
15 & y & y & HN, CH$_3$\\
16 & y & y& O, NH$_2$, CH$_4$\\
17 & y & y & HO, NH$_3$\\
18 & y & y &  H$_2$O\\
25 & - & y &  HC$_2$ \\
26 & y & y & CN\\
27 & y & y & HCN\\
28 & y & y &  CO, N$_2$\\
29 & y & y & HCO, $^{13}$CO, CH$_3$N\\
30 & y & y & H$_2$CO, CH$_3$NH, NO\\
31 & y & y & CH$_3$O,CH$_3$NH$_2$\\
32 & - & y &  CH$_3$OH, O$_2$ \\
33 & - & y &  NH$_2$OH\\
34 & - & y &  H$_2$O$_2$ \\
36 & - & y &  ? \\
38 & - & y &  ? \\
39 & y & y & CHCN \\
40  & - & y &  CH$_2$CN \\
41 & y & y & CH$_3$CN, CH$_2$CNH, CHCO\\
42 & y & y & NCO, CH$_2$CO\\
43 & y & y & HNCO, CH$_3$CO\\
44 & y & y &CO$_2$, N$_2$O, CONH$_2$, CH$_3$CHO\\
45 & y & y & $^{13}$CO$_2$, HCONH$_2$, HCOO\\
46 & y & y & HCOOH   \\
47 & - & y &   HNO$_2$?\\
54 & - & y &   ?\\
55 & y & y & CHNCO   \\
56 & y & y & CH$_2$NCO   \\
57 & y & y & CH$_3$NCO  \\
58 & - & y &  HCOHCO \\
59 & - & y &  C$_2$H$_5$NO \\
60 & - & y & HCOOCH$_3$, NH$_2$CONH$_2$ \\
61 & - & y & CH$_3$NO$_2$  \\
70 & - & y &   ?\\
71 & - & y &   CH$_3$CH$_2$NCO \\
73 & - & y &   HCONHHCO, HCOCONH$_2$\\
75 & - & y &   NH$_2$CH$_2$COOH\\
88 & - & y  &  NH$_2$COCONH$_2$, NH$_2$CONHCHO \\
175 & - & y &   C$_6$H$_9$NO$_5$ \\
\enddata
\end{deluxetable}
\subsection{Warm Up: Infrared and Mass Spectra}
\begin{figure*}[ht!]
\centering
\includegraphics[width=17.cm]{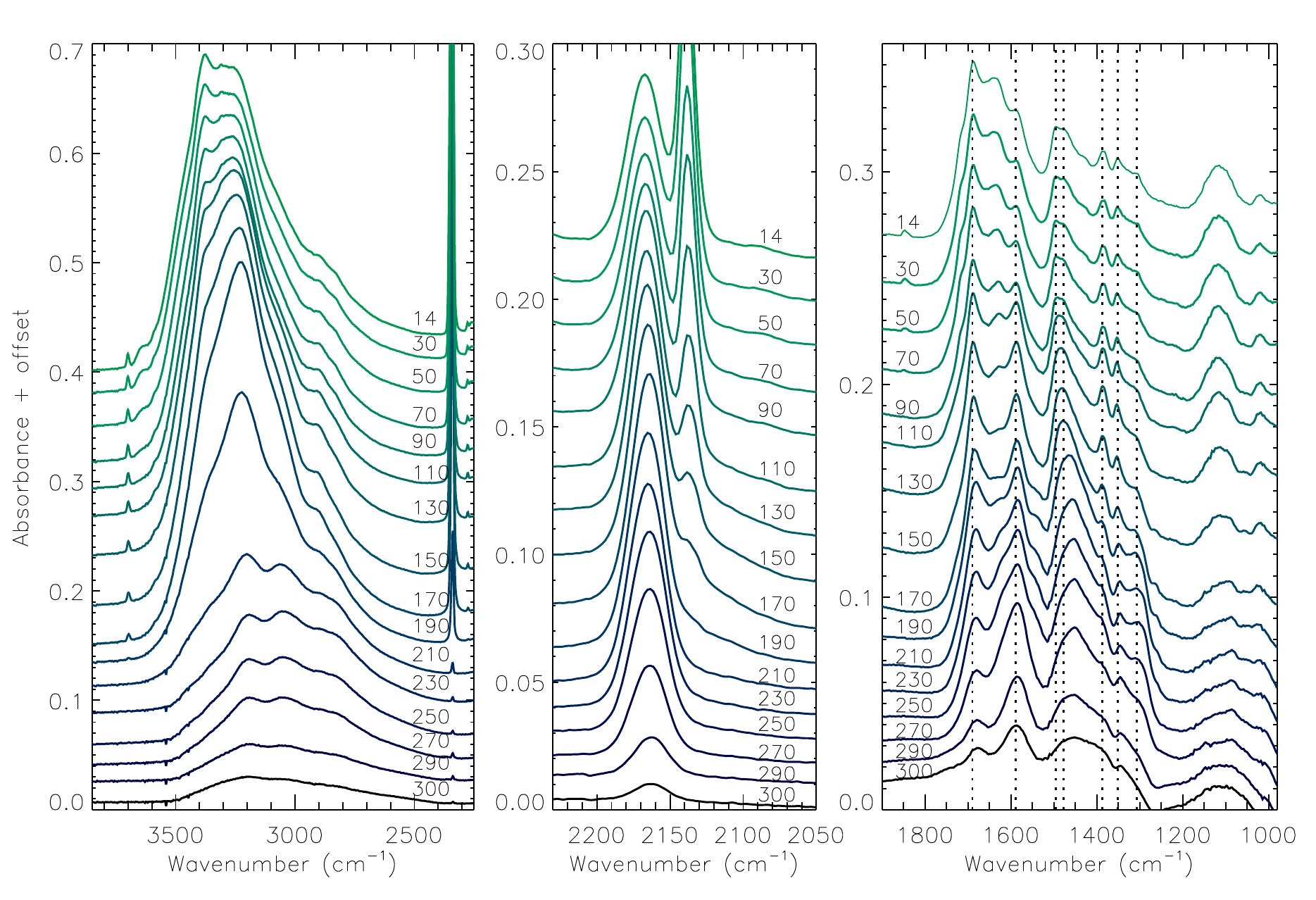} 
\caption{Infrared spectra during the warm up phase from 14 to 300 K for three  spectral ranges. The spectra are taken with 4~cm$^{-1}$ resolution and have been offset for clarity.}
\label{f6}
\end{figure*}

\begin{figure*}
\centering
\includegraphics[width=17cm]{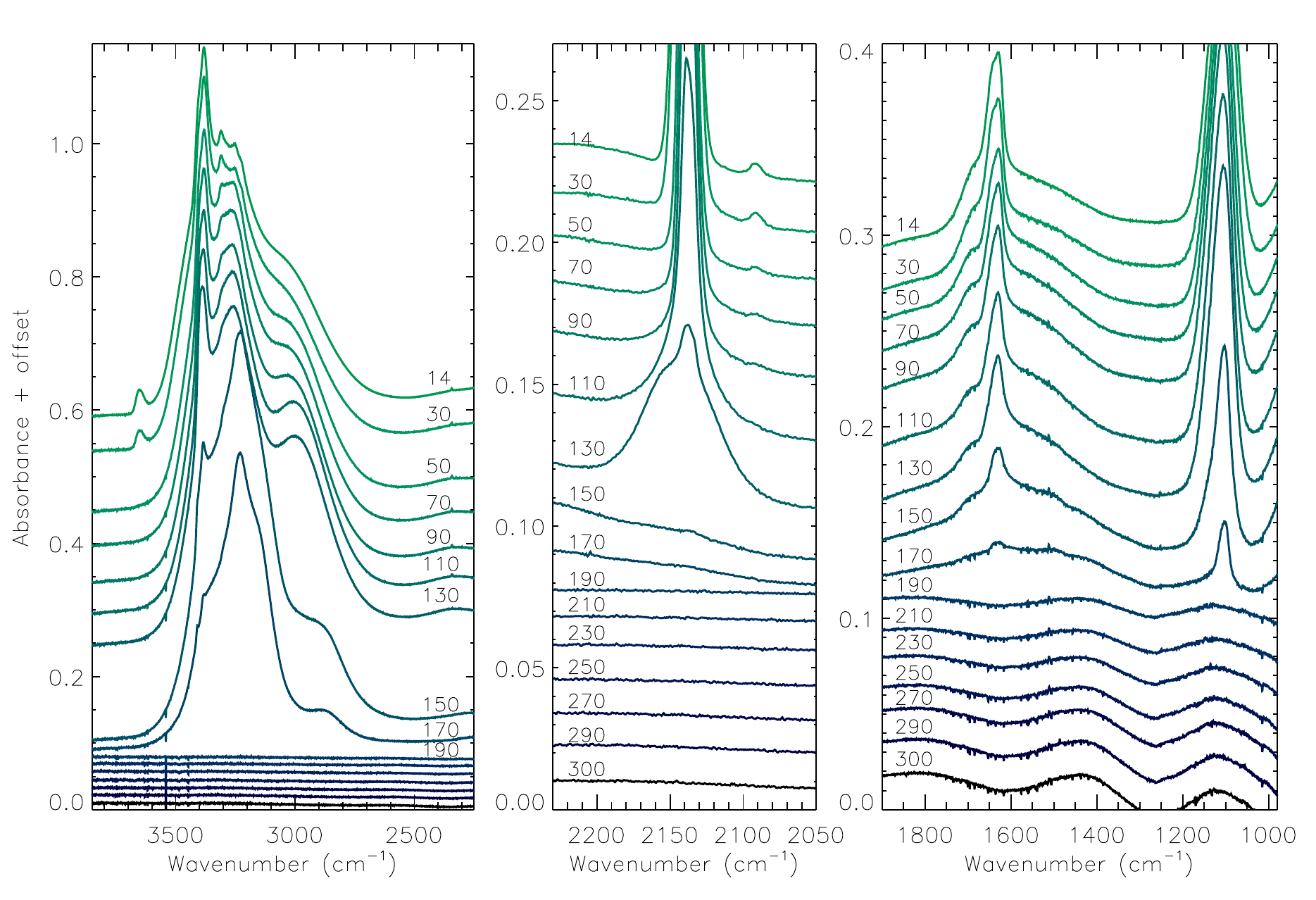} 
\caption{As in Figure~\ref{f6} but for the blank experiment.}
\label{f7}
\end{figure*}

At the end of the irradiation the ice has been warmed up at a rate of 2~K/min. In Figure~\ref{f6} are reported the infrared spectra of the ice at different temperatures during this phase. As comparison we report similar spectra  for the blank experiment in Figure~\ref{f7}.

The desorption temperatures of species in the mixture are generally higher than those in single component ices \citep{Mar14}. CO band at 2145 cm$^{-1}$ disappears completely between 170 and 190 K in the irradiated sample, and between 130 and 150 K in the blank. In a thick ($\approx 10^{18}$~cm$^{-2}$) pure CO ice such infrared feature disappears between 40 and 50~K. As listed in Table~\ref{t2} many masses were detected by the QMS during this phase.

\begin{figure}[ht!]
\centering
\includegraphics[width=9.cm]{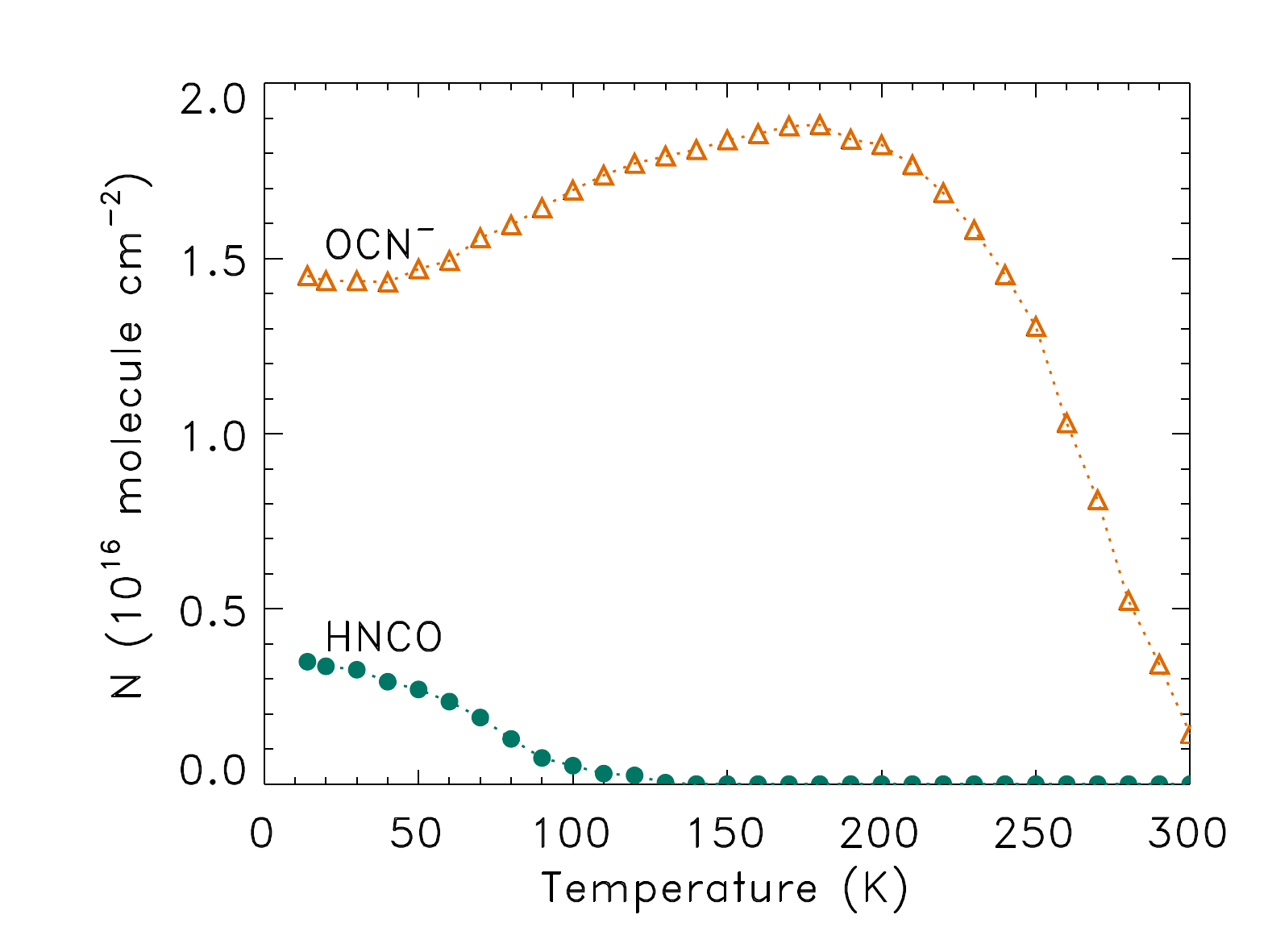} 
\caption{OCN$^-$ (yellow triangles) and HNCO (green bullets) column densities as function of the ice temperature during the warm up.}
\label{f8}
\end{figure}

\begin{figure}[ht!]
\centering
\includegraphics[width=9.5cm]{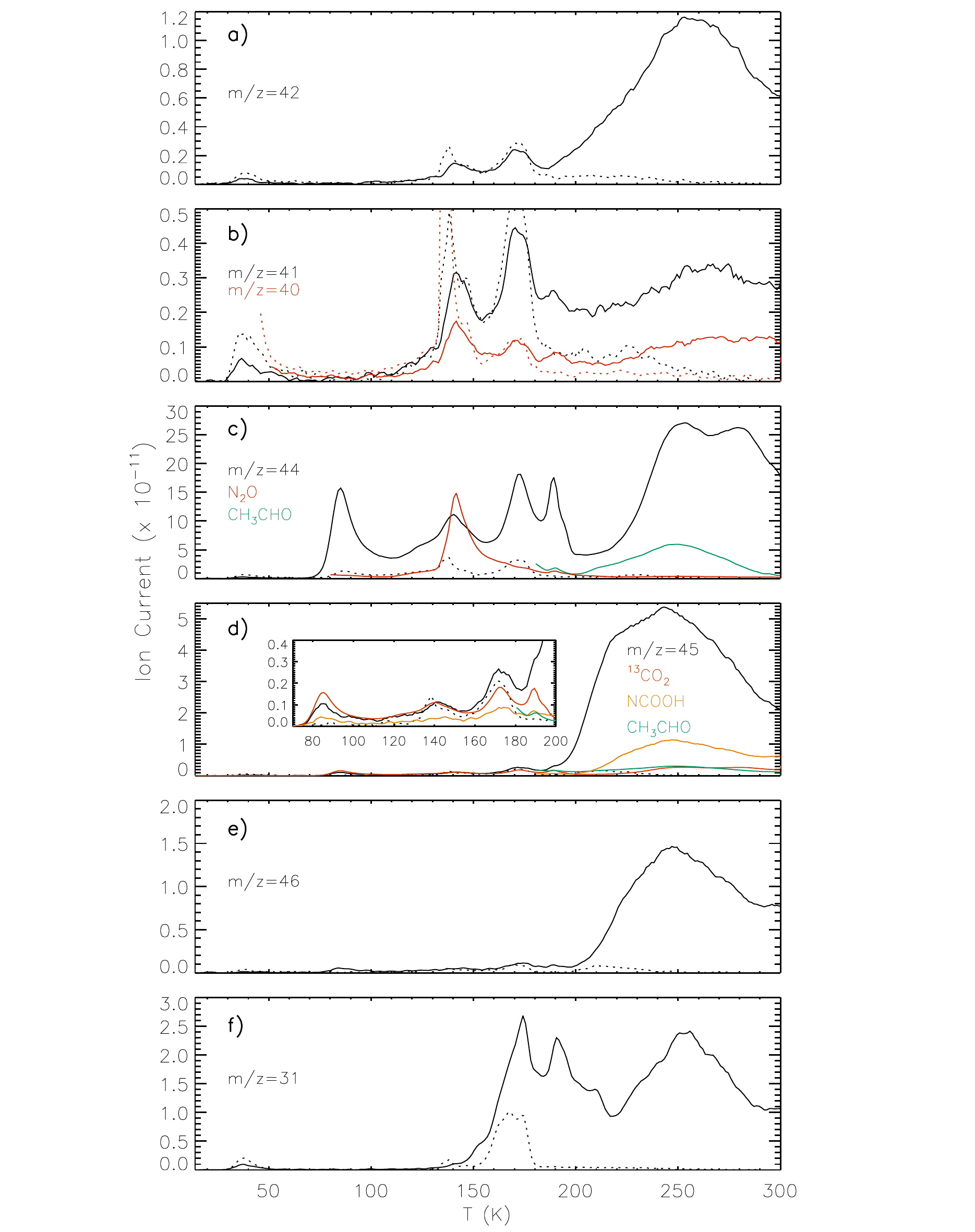} 
\caption{ Ion current vs temperature during the warm up of the ice for $m/z= 42$, 44, 45, 46 and 31. The solid and dotted lines are from the irradiated and blank experiments, respectively. In the second and third panels from the top are shown in colors species that contribute to $m/z=44$ and 45, respectively.}
\label{f9}
\end{figure}

OCN$^-$ at 2168 cm$^{-1}$ is still present at 300 K, its column density  as function of the ice temperature is shown in Figure~\ref{f8}. The column density of the HNCO in the same figure is, as in Figure~\ref{f5}, the sum of the 2241 and 2261 cm$^{-1}$ bands. Although band strengths may change with temperature (e.g. \citealt{Luna2018}), no measurements  on either OCN$^-$  and HNCO are available. Data reported in the Figure~\ref{f8} are thus obtained taken the band strengths constant.
The band strengths at 15 K of 1.3 $\times$ 10$^{-16}$~cm~molecule$^{-1}$ for OCN$^-$ \citep{Bro04}, and 7.8 $\times$ 10$^{-17}$~cm~molecule$^{-1}$ for HNCO \citep{Bro04} have been used. During the warm up OCN$^-$ increases from 50 up to 170 K, then starts decreasing. Such increase could be justified by a further production of OCN$^-$ through the reaction $\rm HNCO + NH_3 \rightarrow OCN^- + NH_4^+$. In agreement with the infrared spectra, starting from  170 K the ion current of mass $m/z=42$ increases reaching a broad peak between 240 and 270 K, with a tail extending up to 300 K (see Figure~\ref{f9}).

As mentioned in the previous section the infrared feature at 2260~cm$^{-1}$ is compatible with acetonitrile CH$_3$CN. During the warm up $m/z=41$ associated to such species or its isomer CH$_2$CNH is detected along with $m/z=40$, see second panel from the top in Figure~\ref{f9}. $m/z=40$ is the largest fragments after $m/z=41$ for both CH$_3$CN and CH$_2$CNH. Above $\sim$ 200 K the profiles of such masses are very similar supporting the identification of the species.

\begin{figure}[ht!]
\centering
\includegraphics[width=9.cm]{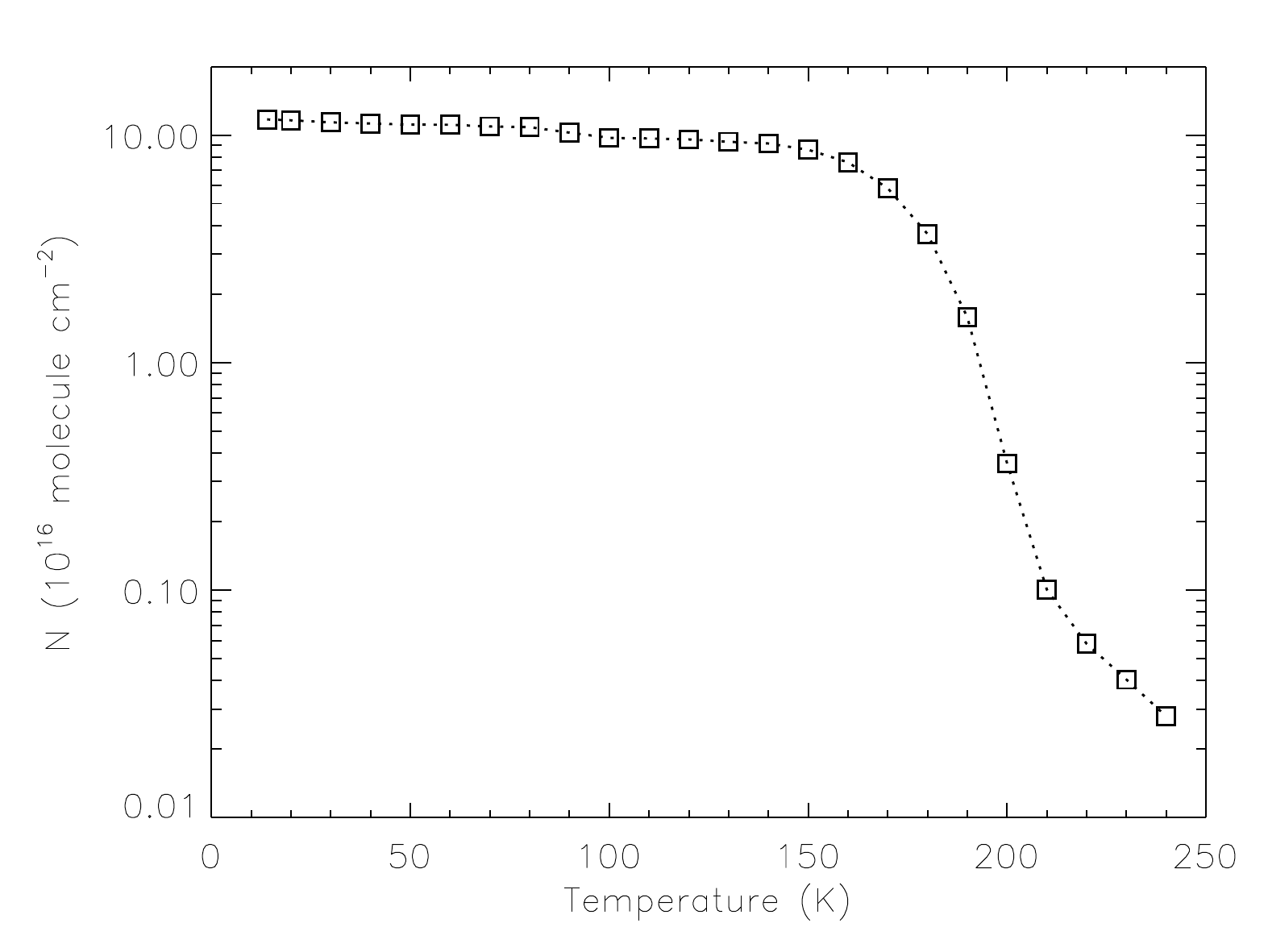} 
\caption{CO$_2$ column density as function of the ice temperature during the warm up.}
\label{f10}
\end{figure}

The column density of CO$_2$ (2342 cm$^{-1}$) shown in Figure~\ref{f10} as function of the temperature during the warm up phase indicates that about 90\% of CO$_2$ desorbs between 150 and 190 K (see also Figure~\ref{f6}). Although the desorption peak of a pure CO$_2$ is around 85~K, a fraction of this molecule may be retained in the ice matrix, and co-desorbed with others abundant species such as water. In Figure~\ref{f9} the mass spectrum of $m/z=44$ shows a series of desorption peaks between 80 and 300~K. While the peaks between  80 and 200~K are in agreement with desorption of CO$_2$, as shown by the infrared spectra in Figure~\ref{f6} and the column density in Figure~\ref{f10}, the broad and intense peak between 220 and 300~K suggests a contribution to $m/z=44$ from fragmentation of larger compounds in the ice.

An upper limit to the contribution of N$_2$O to $m/z=44$ can be estimated using $m/z=14$ a fragment of either NH$_3$ and N$_2$. Assuming that $m/z=14$ is only associated to the fragmentation of N$_2$O, we obtain a negligible contribution of N$_2$O to $m/z=44$ (see red curve in the third panel from the top of Figure~\ref{f9}). 

One of the carrier of the band at 1352 cm$^{-1}$ is CH$_3$CHO ($m/z=44$). $m/z=44$ of this species is $\sim$ 83\% of its main mass $m/z=29$. Even assuming that $m/z=29$ at high temperature is only related to CH$_3$CHO, an upper limit to its contribution is given by the green curve shown in panel c) of Figure~\ref{f9}.

Several bands in the range in $1000-1800$~cm$^{-1}$ of the infrared spectra survive up to room temperature. In particular, the bands at 1690~cm$^{-1}$ of formamide is clearly present in the infrared spectra at 300~K. The associate mass $m/z = 45$ in panel d) of Figure~\ref{f9} shows a series of peaks within the temperature range. The red curve in the panel is the contribution of $^{13}$CO$_2$ to $m/z=45$, computed as the 10\% of $m/z=44$.  Assuming that CO$_2$ is the only carrier of $m/z = 44$, its isotopologue $^{13}$CO$_2$ can at most justify $m/z=45$ only below 200 K. Pure formamide in thick ice has a desorption peak at 220~K that well fit with the broad feature between 200 and 300 K. In this range acetic acid (HCOOH - yellow curve) and acetaldehyde (CH$_3$CHO - green curve) may also contribute to this mass.

\begin{figure}
\centering
\includegraphics[width=8.cm]{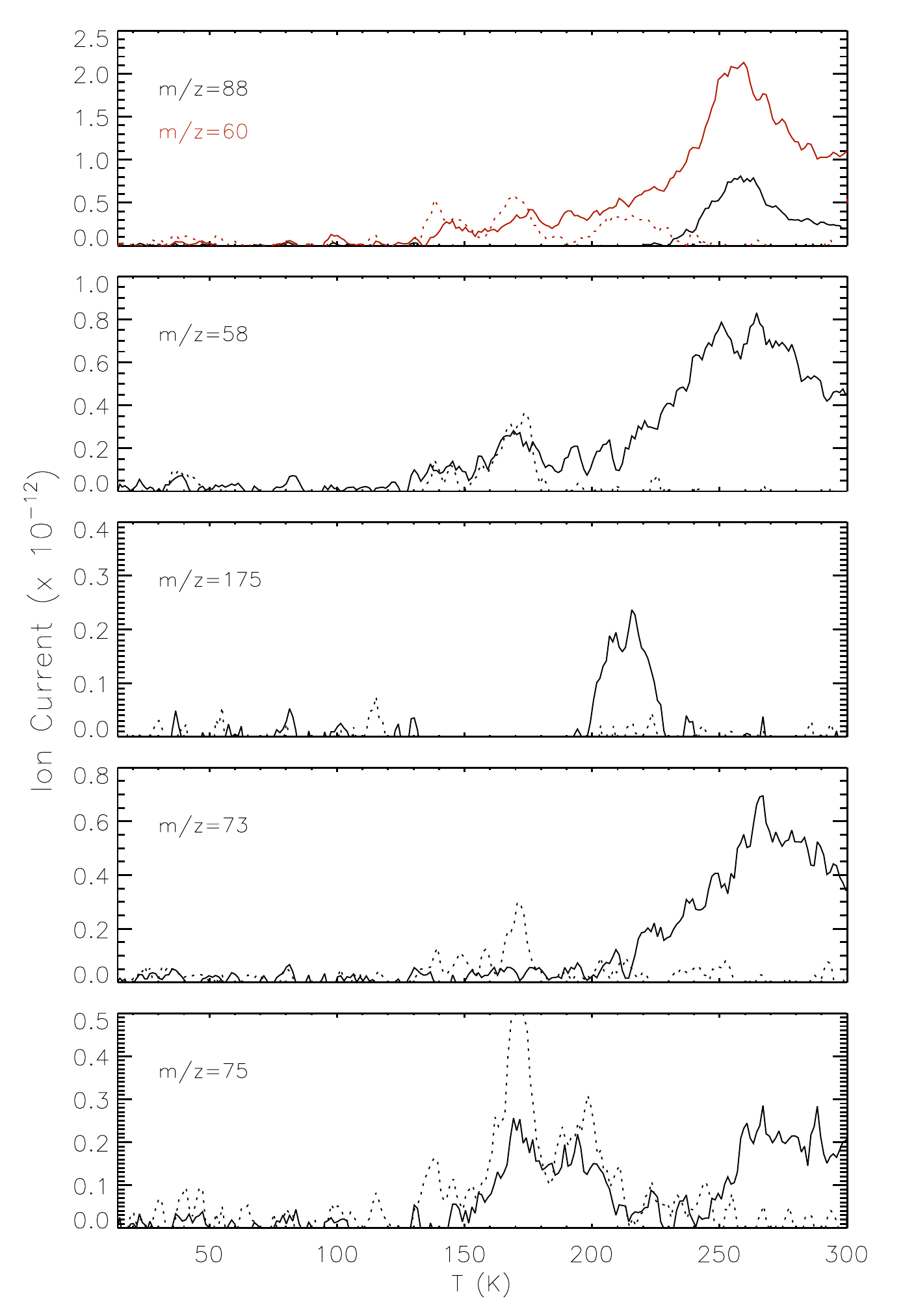} 
\caption{Ion current vs temperature during the warm up of the ice for $m/z= 88$, 60, 58, 175 and 75. The solid and dotted lines are from the irradiated and blank experiments, respectively.}
\label{f11}
\end{figure}

The band at 1590 cm$^{-1}$ is associated to three different species: HCOO$^-$ that most probably desorbs after recombination with H as $m/z=46$, see panel e); CH$_3$NH$_2$ ($m/z=31$), see panel f); NH$_3^+$CH$_2$COO$^-$ ($m/z= 75$). This band is still present in the infrared spectra at 300 K (Figure~\ref{f7}). These masses shown in Figure~\ref{f9} and that of NH$_3^+$CH$_2$COO$^-$ in Figure~\ref{f11} have desorption peaks extending up to 300 K. 

The QMS is much more sensitive than the infrared spectrometer and some of the detected masses during the warm up have no clear features in the infrared spectra. Complex organic molecules produced in the experiment have generally low abundances, share similar fragments and are therefore more challenging to identify. Masses such as $m/z=28$, 29, 30, 31, 32 and 44 are common to many species and their intensities at high temperature require many contributors. For such species to be identified, we require the presence of the expected main mass fragments, and the existence of a consistent reaction channel in the ice.

Polymerization of two HCO$^\cdot$ radicals in the mixture will bring to glyoxal species (HCOHCO, $m/z=58$)
\begin{equation}
\rm
2 HC^{\cdot} O \longrightarrow {\small\chemfig{(-[3,0.5]H)(=[5,0.5]O)-[0,0.5](=[1,0.5]O)-[7,0.5]H}}
\end{equation}
During warm-up, $m/z = 58$ show a desorption peak around 260~K (see Figure~\ref{f11}). Other fragments of such species, i.e. $m/z=29$, 31 and 30 are detected co-desorbing with $m/z=58$, support the presence of glyoxal.

Vacuum ultraviolet irradiation experiments of CH$_4$:HNCO \citep{Lig18} produce the desorption of a fragment with $m/z=73$ at $\approx$ 220~K, that have been assigned to several candidates, the most probable being propionamide, CH$_3$CH$_2$C(O)NH$_2$. 

 In our experiments $m/z = 73$ desorbs at higher temperatures, $\approx$ 270 K. Considering the different ice composition and thus, the different first generation products such as H$_2$CO and HCONH$_2$, $m/z = 73$ is readily explained by further energetic processing yielding a second generation of radicals
\begin{equation}
\rm
HCO^\cdot + N^{\cdot}HCHO \longrightarrow {\small\chemfig{(=[2,0.5]O)(-[5,0.5]H)-[7,0.5]NH-[1,0.5](=[2,0.5]O)-[7,0.5]H}}
\end{equation}
\begin{equation}
\rm
HCO^\cdot + NH_2C^{\cdot}O \longrightarrow {\small\chemfig{(-[3,0.5]H)(=[5,0.5]O)-[0,0.5](=[1,0.5]O)-[7,0.5]NH_2}}
\end{equation}
Radical reactions of this species are among the most important routes in producing new molecules of astrobiological interest. 

As mentioned in Section~\ref{prod} and Table~\ref{t1} several infrared features are compatible with glycine or other isomers. The thermal desorption of $m/z=75$ further support the infrared identification. As shown in Figure~\ref{f11} this mass although very weak shows a desorption from 250 to 300K. Its main fragment $m/z=30$, presenting a desorption tail extending up to room temperature, supports this assignment.

As the chemical complexity induced by energetic processing increases, molecules with $m/z=88$ can be produced by reactions involving second generation radicals. The reaction between two formamide radicals bring to ethanediamide, NH$_2$COCONH$_2$
\begin{equation}
\rm
2 NH_2C^{\cdot}O \longrightarrow 
{\small\chemfig{(-[3,0.5]NH_2)(=[5,0.5]O)-[0,0.5](=[1,0.5]O)-[7,0.5]NH_2
}}
\end{equation}
Ethanediamide desorption is supported by detection of the fragment NH$_2$CONH$_2$ ($m/z=60$). As Figure~\ref{f11} shows the two masses have similar profile. Since the ratio $(m/z=60)/(m/z=88)$ is much larger than the value of 1.2 expected for ethanediamide, contributions from other species (e.g., urea) are required. The infrared bands of urea could be blended with the features around 1150 and 1500 cm$^{-1}$. Reaction of urea radical with HCO produces formylurea (NH$_2$CONHCHO, $m/z=88$),
\begin{equation}
{\small \chemfig{C^\cdot(=[2,0.5]O)(-[5,0.5]H)} + 
\chemfig{(=[2,0.5]O)(-[5,0.5]NH_2)(-[7,0.5]N{^\cdot}H)}} 
\longrightarrow {\small\chemfig{(=[2,0.5]O)(-[5,0.5]H)-[7,0.5]NH-[1,0.5 ](=[2,0.5]O)-[7,0.5]NH_2}}
\end{equation}
 
Methyl-formate, HCOOCH$_3$, is another candidate for $m/z=60$. This species as shown in Table~\ref{t1} it could contribute to the 1717~cm$^{-1}$ feature. Its main fragments $m/z=31$, 32 and 29 are observed desorbing at the same temperature.  

The largest mass detected during the warm up is $m/z=175$. Such a large mass is difficult to identify. A possible candidate for $m/z=175$ could be N-acetyl-L-aspartic acid. Such species implies the presence of aspartic acid and other amino acids like alanine. Aspartic acid may in fact be present given the detection of his main fragments $m/z=88$, 70 and 43. Alanine has as main fragment $m/z=44$, a mass common to many species. A more reliable identification of molecules having such large mass requires the use of other analytical techniques, and/or the  characterization of the residue, that is beyond the scope of this work.

Table~\ref{t2} summarize the species identified as carriers of the observed mass spectra.

\subsection{The Refractory Residue}
\begin{figure}
\centering
\includegraphics[width=9.cm]{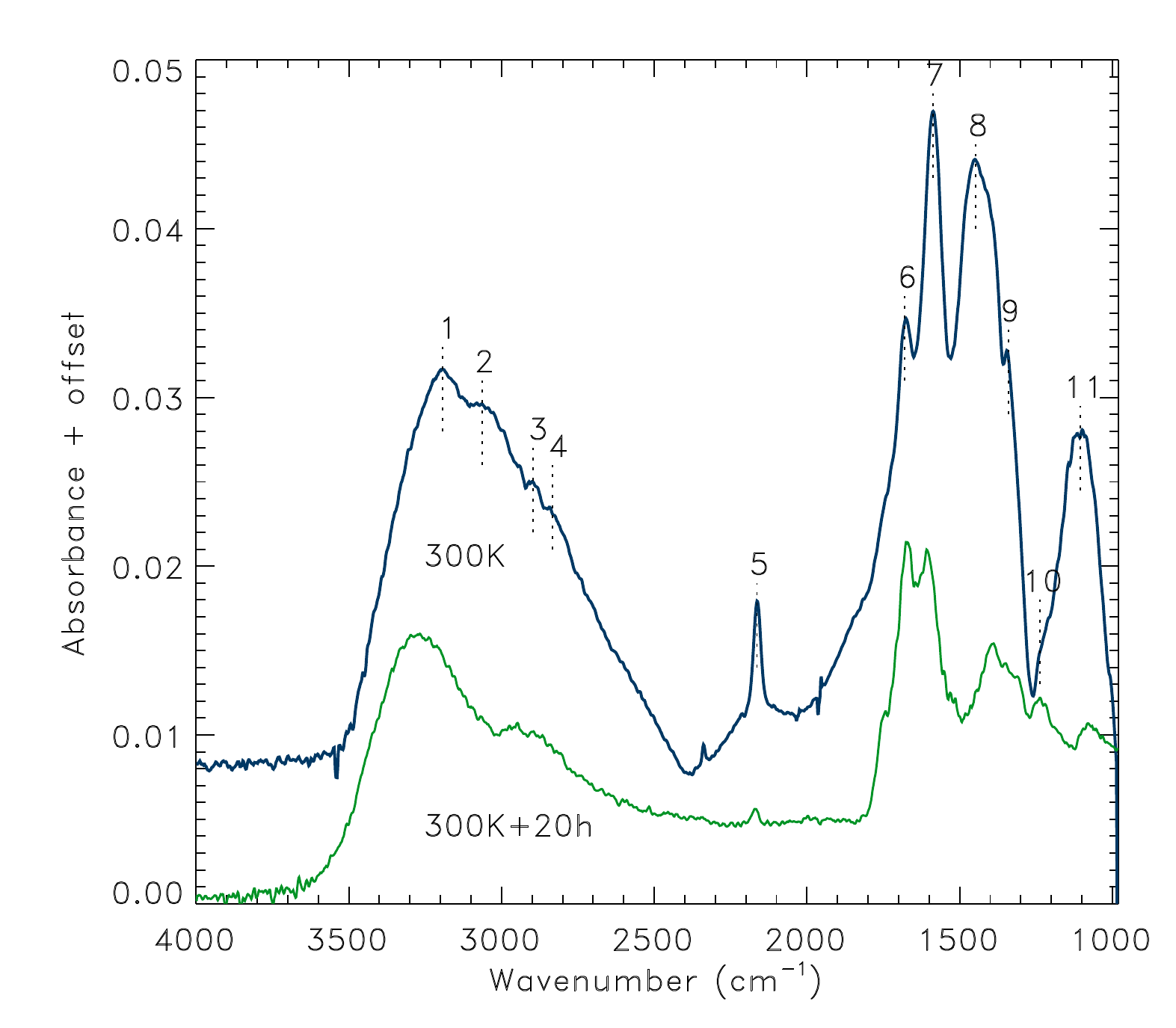} 
\caption{Infrared spectra of the residue at the end of the warm up at 300 K (blue curve) and after 20 h (green curve) in the vacuum chamber. The dotted lines mark the main features. The curves are shifted for clarity.}
\label{fres}
\end{figure}

Infrared spectra of the residue are shown in Figure~\ref{fres}. The two spectra were obtained at 300 K, the first just after the warm-up and the second one day after keeping the window for 20 h  in the vacuum chamber. Although the spectrum after 20~h is significantly reduced, in the range $1000-2400$~cm$^{-1}$ all the features are still present.  The broad band around 3000 cm$^{-1}$ is shared by species containing the hydroxyl functional group, such as alcohols and carboxylic acids frequently present in processed ice residues. After 20~h, this band displays a clear tail that extends beyond 2500 cm$^{-1}$, suggesting the presence of carboxylic acids. Indeed, alcohols are expected to desorb gradually at room temperature, which likely accounts for the decrease of the band around 3000 cm$^{-1}$ and the band peaking around 1100 cm$^{-1}$. The feature at 2165 cm$^{-1}$ typical of the C=N stretch in OCN$^-$ still present after 20~h.

The COO$^-$ feature at 1588~cm$^{-1}$ typical of carboxylic acid salts is significantly reduced after 20~h, as compared to the 1679 cm$^{-1}$. The C=O stretching  at  1679 cm$^{-1}$ suggests the presence of amides. 

Hexamethylenetetramine (HMT) in the spectrum corresponds to the bands at 2897, 2832, and 1238 cm$^{-1}$. The main HMT band peaks around 1007~cm$^{-1}$, in this spectrum it would be a small narrow band superposed on the larger band that appears at this wavenumber. In ultraviolet experiments with ice of the present composition, HMT is very minor species, that makes sense because HMT has 6~CH$_2$ groups, and CO is not a good precursor; the presence of methanol would be much more favorable for its formation. In addition, the formaldehyde precursor of HMT is more efficently formed when methanol is included in the ice. Furthermore, HMT forms at room temperature, and it would not be expected in the fresh residue. The feature at $\sim 1105$~cm$^{-1}$ that shifts at lower wavenumber in the spectrum after 20 h is associated to HOCH$_2$COO$^−$ \citep{Mu03}. The bands identified in the infrared spectrum of the residue are in Table~\ref{t3}.
\begin{deluxetable}{llcl}
\tabletypesize{\scriptsize} 
\tablecaption{Identification of the features in the residue \label{t3}}
\tablewidth{0pt}
\tablehead{Label & Band &  Assignment & Ref.\\
& (cm$^{-1}$) &  & }
\startdata
1 & 3192 & NH$_4^+$ & 1 \\
2 & 3063 & ?? &  \\
3 & 2897 & HMT?& 1\\
4 & 2832 & HMT,NH$_4^+$& 1\\
5 & 2165 & OCN$^-$  &  2 \\
6 & 1679 & Amides & 1,2\\
7 & 1588 & COO$^-$ str. in carboxylic acid salts & 1,2\\
8 & 1447  & NH$_4^+$ & 1 \\
9 & 1341  & COO$^-$ str. in carboxylic acid salts & 1,2 \\
10 & 1238 & HMT & 1\\
11 & 1105 & HOCH$_2$COO$^−$ & 1 \\
\enddata
\tablecomments{(1) \citet{Mu03}; (2) \citet{Pil10}}
\end{deluxetable} 

\begin{deluxetable}{ccc}
\tabletypesize{\scriptsize} 
\tablecaption{Photo-desorption Yields\label{t4}}
\tablewidth{0pt}
\tablehead{m/z & Molecule &  molecules/photon }
\startdata
31 & CH$_3$NH$_2$ & 1.94 $\times$ 10$^{-5}$\\
41 & CH$_3$CN     & 3.62 $\times$ 10$^{-5}$\\
43 & HNCO         & 3.17 $\times$ 10$^{-5}$\\
57 & CH$_3$NCO    & 2.03 $\times$ 10$^{-5}$
\enddata
\label{pd}
\end{deluxetable} 

\newpage
\section{Conclusions and Astrophysical Implications}\label{disc}
We have studied the chemical evolution induced by soft X-rays in the range $250 - 1250$~eV in a mixture containing nitrogen, H$_2$O:CO:NH$_3$. Among the products of the irradiation there are many N-bearing molecules such as e.g., OCN$^-$ (2168 cm$^{-1}$), HNCO and CH$_3$CN (2261 cm$^{-1}$), HCONH$_2$ (1690 cm$^{-1}$), and NH$_4^+$ (1478 cm$^{-1}$). Several features of the irradiated mixture are compatible with the presence of glycine or its isomers (see e.g., \citealt{Ob16}). The products detected in the infrared spectra are common to other irradiation experiments of similar mixtures with ultraviolet \citep{Hu00} and energetic particles \citep{Hu00,Pil10}. 

Detection of several masses in gas-phase either during the irradiation and the warm up revealed a number of complex organics with not clear features in the infrared spectra of the irradiated ice. Photo- and thermal-desorptions are of great interest as they offer the possibility that complex organics formed in solid phase can enrich the gas-phase molecular distribution.

Acetonitrile CH$_3$CN detected through the 2260 cm$^{-1}$ infrared feature and $m/z=41$ has been observed in Sgr~B2 and Sgr~A \citep{Sol71} while its isomer CH$_2$CN (compatible with 2038 cm$^{-1}$) in Sgr~B2 \citep{Lo06}.  More recently methyl cyanide has been observed in a protoplanetary disc around MWC480 by \citet{Ob15}.

Formamide HCONH$_2$ is also produced and its mass ($m/z=45$) is among the most intense during thermal desorption. Formamide with its peptide moiety may be a building block for species of astrobiological interest such as sugars and amino-acids. This species has been identified in hot cores, and external galaxies (cf. \citealt{Mcg18}), and also around the SVS13-A protostar by \citet{Bia19}. These authors also detected acetaldehyde (CH$_3$CHO), methanol (CH$_3$OH), and methyl formate (HCOOCH$_3$), all of them produced in the present experiments, and revealed through QMS in the gas-phase either during the irradiation or warm up (see Table~\ref{t2}).

Another species detected in the gas-phase during the experiments is methyl isocyanate (CH$_3$NCO), $m/z=57$, itself a molecule of prebiotic relevance. This species has been recently detected in protostar environments \citep{Lig17,Mar17}, the Orion cloud \citep{Cer16}, and in comets \citep{Goe15}. \citet{Lig17} estimated the column density of CH$_3$NCO towards the binary low-mass protostar IRAS 16293-2422 to be $\sim (3-4) \times 10^{15}$ and $\sim (6-9) \times 10^{15}$~cm$^{-2}$ are obtained towards sources B and A, respectively. A similar value, $4 \times 10^{15}$~cm$^{-2}$, has been derived towards IRAS~16293-2422~B by \citet{Mar17}. The fractional abundances with respect to hydrogen nuclei results $\la 1.4 \times 10^{-10}$ \citep{Mar17}.

Following the QMS quantification and calibration by \citet{Mar15}, we estimated the photo-desorption yields in molecules~ph$^{-1}$ for the molecules listed in Table~\ref{pd}. The quantitative calibration of QMS for the present facility has been detailed in \citet{NS19}. We then exploit the chemical model of protoplanetary disc around classical T Tauri stars put forward by \citet{W12}. These authors investigated an axisymmetric region surrounding a star with half the mass and twice the radius of the Sun. The effective temperature is 4000~K, and the X-ray luminosity, $1 \times 10^{30}$~erg~s$^{-1}$, is taken constant throughout the disc evolution. The stellar X-ray spectrum used in the disc modelling \citep{N07} covers energies from 0.1 to 10 keV with a dominant broad and intense peak in the range between 200 and 900~eV, similar in shape to the spectrum used in our experiments (see Figure~\ref{f1}).

The gas-phase fractional density of a particular species $i$ produced by photo-desorption is given by
\begin{equation}
f_i = Y_i \langle A_{\rm g} \rangle F_{\rm X}\tau_{\rm X}
\label{tau}
\end{equation}
where $Y_i$ is the photo-desorption yield reported in Table~\ref{pd}, $\langle A_{\rm g} \rangle \sim 1 \times 10^{-22}$~cm$^2$ the average dust geometrical cross-section per hydrogen nucleus, $F_{\rm X}$ the local X-ray flux in photons~cm$^{-2}$~s$^{-1}$, and $\tau_{\rm X}$ the irradiation time.

Methyl isocyanate emission has been detected towards IRAS~16293-2422 on Solar System scales, i.e. within $\sim 60$~au from the prostar \citep{Lig17}.  At radial distances between 10 and 50~au, and within 50~au of altitude from the plane of the disc, dust temperatures are between 20 and 40~K,  consistent with a long residence time of mixed water ices. The X-ray flux results $F_{\rm X} \sim 10^{-4} - 10^{-2}$~ergs~cm$^{-2}$~s$^{-1}$ \citep{W12}. Assuming an average photon energy of 600 eV, we obtain $F_{\rm X} = 10^5 -10^7$~photons~cm$^{-2}$~s$^{-1}$. Using equation~(\ref{tau}), we derive an irradiation time $\tau_{\rm X} \ga 200 - 20000$~yr. Such times are compatible with the disc dynamical age of 40000~yr reported in \citet{Mar17}, and with other estimates for this source \citep{Bot14,Ma16}.

On the base of Table~\ref{pd} we may estimate the CH$_3$NCO/HNCO ratio to be $\sim 0.6$, about a factor of 3 bigger than the largest observational estimate $\la 0.25$ \citep{Lig17}. However, we did not expect to find a close agreement with the observational data, as the two species, CH$_3$NCO and HNCO formed on the surface of dust grains, are photo-desorbed and incorporated into the gas phase without any further interaction. Our results show the photo-desorption may be an efficient non-thermal source of such species, and other organic molecules. Thus, around young solar-type stars, known to emit many more X-rays than the present-day Sun, X-ray irradiation has a potential role in prebiotic chemistry. 

In conclusion, the present results show that X-ray irradiation of a water mixed (H$_2$O:CO:NH$_3$) ice may originate a chemistry rich in organic compounds, most of them likely to be desorbed to the gas-phase via photo and/or thermal desorption. Such species have been detected in many astrophysical environments and in particular in circumstellar regions. In this latter case, X-rays can permeate the disc and reach deeper regions, where less energetic radiation such as ultraviolet is inhibited, triggering a solid phase chemistry that through  photo-desorption can enrich  dilute medium of complex organics. Future experiments will be aimed to a more detailed identification of the species including the solid and gas-phase production yields in order to provide a more accurate comparison with observations.

\section{Acknowledgments}
We acknowledge the NSRRC general staff for running the synchrotron 
radiation facility. We also thank Dr. T.-W.
Pi, the spokesperson of BL08B in NSRRC.

This work has been supported by the project PRIN-INAF 2016 The Cradle of Life - GENESIS-SKA (General Conditions in Early Planetary Systems for the rise of life with SKA). We also acknowledge support from INAF through the "Progetto Premiale: A Way to Other Worlds" of the Italian Ministry of Education. One of us, G.M.M.C., was financed by the Spanish MINECO-FEDER under projects AYA2014-60585-P and AYA2017-85322-R, and the MOST grants MOST 107-2112-M-008-016-MY3 (Y.-J.C.), Taiwan.


\begin{thebibliography}{}
\bibitem[Aikawa et al.(2012)]{Aik12} Aikawa, Y. Kamuro, D., Sakon, I., et al. 2012, \aap, 538, 57
\bibitem[Andrade et al.(2010)]{A10} Andrade, D. P. P., Rocco, M. L. M., \& Boechat-Roberty, H. M. 2010, \mnras, 409, 1289
\bibitem[Bernstein et al.(1995)]{Ber95} Bernstein, M. P., Sandford, S. A., Allamandola, L. J., Chang, S., \& Scharberg, M. A. 1995, \apj, 454, 327
\bibitem[Benndorf et al.(1999)]{Be99}Benndor, M., Westerveld, W.B., van Eck, J., van der Weg, J. \& Heideman, H.G.M. 1999, J. Phys. B: At. Mol. Opt. Phys., 32, 2503
\bibitem[Bernstein et al.(2002)]{Ber02} Bernstein, M. P., Dworkin, J. P., Sandford, S. A., Cooper, G. W., \& Allamandola, L. J. 2002, \nat, 416, 401
\bibitem[Bianchi et al.(2019)]{Bia19} Bianchi, E., Codella, C., Ceccarelli, C., et al. 2019, \mnras, 483, 1850
\bibitem[Bisschop et al.(2007)]{Bis07}Bisschop, S. E., Fuchs, G. W., van Dishoeck, E. F., Linnartz, H. 2007, \aap, 464, 1061
\bibitem[Boogert et al.(2015)]{Bo15} Boogert, A. C. A., Gerakines, P. A., \& Whittet, D. C. B. 2015, ARA\&A, 53, 541
\bibitem[Bottinelli et al.(2014)]{Bot14} Bottinelli S., Wakelam V., Caux E., Vastel C., Aikawa Y., \& Ceccarelli C. 2014, \mnras, 441, 1964
\bibitem[Brucato et al.(2006)]{Br06} Brucato, J. R.,  Baratta, G. A., \& Strazzulla, G. 2006, \aap, 455, 395
\bibitem[Cernicharo et al.(2016)]{Cer16} Cernicharo, J., Kisiel, Z., Tercero, B., et al. 2016, \aap, 587, L4
\bibitem[Chen et al.(2007)]{Ch07} Chen, Y.-J., Nuevo, M., Hsieh, J.-M., et al. 2007, \aap, 464, 253
\bibitem[Chen et al.(2013)]{Ch13} Chen, Y.-J., Ciaravella, A., Mu\~{n}oz Caro, G. M., et al. 2013, \apj, 778, 162
\bibitem[Chen et al.(2014)]{Ch14} Chen, Y.-J. , Chuang, K.-J., Mu\~{n}oz Caro, G.M., Nuevo, M., Chu, C.-C. , Yih T.-S. , Ip, W.-H. \& Wu C.-Y. R., 2014, \apj, 781, 15
\bibitem[Chen et al.(2015)]{Che15} Chen, H.-F., Liu, M.-C., Chen, S.-C., et al. 2015, \apj, 804, 36
\bibitem[Ciaravella et al.(2010)]{C10} Ciaravella, A, Mu\~{no}z Caro, G. M., Jim\'enez-Escobar, A., Cecchi-Pestellini, C., Giarrusso, S., Barbera, M., \& Collura, A. 2010, \apjl, 722, L45
\bibitem[Ciaravella et al.(2012)]{C12} Ciaravella, A, Mu\~{no}z Caro, G. M., Jim\'enez-Escobar, A., Cecchi-Pestellini, C., Candia, R., Giarrusso, S., Barbera, M., \& Collura, A. 2010, \apjl, 746, L1
\bibitem[Ciaravella et al.(2016)]{C16} Ciaravella, A., Chen, Y.-J., Cecchi-Pestellini, C., Jim\'enez-Escobar, A., Mu\~{no}z Caro, G. M., Chuang, K.-J., \& Huang, C.-H. 2016, \apj, 819, 38
\bibitem[de Marcellus et al.(2011)]{dM11} de Marcellus, P., Meinert, C., Nuevo, M., et al. 2011, \apjl, 727, L27
\bibitem[Demyk et al.(1998)]{Dem98} Demyk, K., Dartois, E., d’Hendecourt, L., Jourdain de Muizon, M., Heras, A. M., \& M Breitfellner, M. 1998, \aap, 339, 553 
\bibitem[F\"orstel et al.(2016)]{F16} F\"orstel, M., Maksyutenko, P., Jones, B. M., Sun, B. J., Lee, H. C., Chang, A. H. H., \& Kaiser, R. I. 2016, \apj, 820, 117
\bibitem[Gerakines et al.(1996)]{Ger96} Gerakines, P. A., Schutte, W. A., Ehrenfreund, P. 1996, \aap, 312, 289
\bibitem[Gerakines et al.(2004)]{Ger04} Gerakines, P. A., Moore, M. H., \& Hudson, R. L. 2004, \icarus, 170, 202
\bibitem[Gerakines et al.(2005)]{Ger05} Gerakines, P. A., Bray, J. J., Davis, A., \& Richey, C. R. 2005, \apj, 620, 1140
\bibitem[Goesmann et al.(2015)]{Goe15} Goesmann, F., Rosenbauer, H., Bredeh\"oftet, J. K., et al. 2015, Science, 6247, 349
\bibitem[G\'{o}mez-Zavaglia \& Fausto (2003)]{Go03} G\'{o}mez-Zavaglia, A., \& Fausto, R. 2003, PCCP, 5, 3154
\bibitem[Hagen et al.(1979)]{H79} Hagen, W., Allamandola, L. J., \& Greenberg, J. M. 1979, \apss, 65, 215
\bibitem[Herbst \& van Dishoeck(2009)]{Her09}Herbst, E., \& van Dishoeck, E.F. 2009,\araa, 47, 427
\bibitem[Hinkle et al.(1988)]{Hin88} Hinkle, K. H., Keady, J. J., \& Bernath, P. F. 1988, Science, 241, 1319
\bibitem[Holtom et al.(2005)]{Hol05} Holtom, P. D., Bennett, C. J., Osamura, Y., Mason, N. J., \& Kaiser, R. I. 2005, \apj, 626, 940
\bibitem[Hudson \& Moore(2000)]{Hu00} Hudson, R.L., \& Moore, M.H. 2000, \aap, 357, 787
\bibitem[Islam et al.(2014)]{I14} Islam, F., Baratta, G. A., \& Palumbo, M. E. 2014, \aap, 561, A73	
\bibitem[Jamieson et al.(2005)]{Jam05} Jamieson, C.S., Bennett, C., Mebel, A.M., \& Kaise, R.I. 2005, \apj, 624, 436
\bibitem[Jiang et al.(1975)]{Jia75}Jiang, G. J., Person, W. B., \& Brown, K. G. 1975, \jcp, 64, 1201
\bibitem[Jim\'enez-Escobar et al.(2012)]{J12} Jim\'enez-Escobar, A., Mu\~{no}z Caro, G. M., Ciaravella, A., Cecchi-Pestellini, C., Candia, R., \& Micela, G. 2012, \apjl, 751, L40
\bibitem[Jim\'enez-Escobar et al.(2014)]{Ji14} Jim\'enez-Escobar, A., Giuliano, B. M., Mu\~{no}z Caro, G. M., Cernicharo, J., \&
Marcelino, N. 2014, \apj, 788, 19
\bibitem[Jim\'enez-Escobar et al.(2016)]{Jim16} Jim\'enez-Escobar, A., Chen, Y.-J., Ciaravella, A., Huang, C.-H., Micela, G., Cecchi-Pestellini, C. 2016, \apj, 820, 25
\bibitem[Jim\'enez-Escobar et al.(2018)]{Jim18} Jim\'enez-Escobar, A., Ciaravella, A., Cecchi-Pestellini, C.,  Huang, C.-H., Sie, N. E., Chen, Y.-J., \& Mu\~noz-Caro, G.M. 2018, \apj, 868, 73
\bibitem[Jones et al.(2011)]{J11} Jones, B. M., Bennett, C. J., \& Kaiser, R. I. 2011, \apj, 734, 78
\bibitem[Jones et al.(2014)]{J14} Jones, B. M., Kaiser, R. I., \& Strazzulla, G. 2014, \apj, 781, 85
\bibitem[Ka\v{n}uchov\'a et al.(2016)]{K16} Ka\v{n}uchov\'a, Z., Urso, R. G., Baratta, G. A., Brucato, J. R., Palumbo, M. E., \& Strazzulla, G. 2016, \aap, 585, A155
\bibitem[Ligterink et al.(2017)]{Lig17} Ligterink, N. F. W., Coutens, A., Kofman, V., et al. 2017, \mnras, 469, 2219
\bibitem[Ligterink et al.(2018)]{Lig18} Ligterink, N. F. W., Terwisscha van Scheltinga, J., Taquet, V., Jørgensen, J. K., Cazaux, S., van Dishoeck, E. F., \& Linnartz, H., \mnras, 2018, 48,3628
\bibitem[Loeffler et al.(2005)]{L05} Loeffler, M. J., Baratta, G. A., Palumbo, M. E., Strazzulla, G., \& Baragiola, R. A. 2005, \aap, 435, 587
\bibitem[Lovas et al.(2006)]{Lo06} Lovas, F. J., Hollis, J. M., Remijan, A. J., \& Jewell, P. R. 2006, \apjl, 645, L137

\bibitem[Luna et al.(2018)]{Luna2018} Luna, R.,  Molpeceres, G., Ortigoso, J., Satorre, M.~A., Domingo, M., \&  Mat{\'e}, B. 2018, \aap, 617, A116

\bibitem[Majumdar et al.(2016)]{Ma16} Majumdar L., Gratier P., Vidal T., Wakelam V., Loison J.-C., Hickson K. M., \& Caux E. 2016, \mnras, 458, 2, 1859
\bibitem[Mase et al.(1998)]{Ma98}Mase, K., Nagasono, M., Tanaka, S. \& Urisi, T. 1998 J. Chem. Phys., 108, 6550
\bibitem[Mart\'in-Dom\'enech et al.(2014)]{Mar14}Mart\'in-Dom\'enech, R., Mu\~noz-Caro, G.M., Bueno, J., et al. 2014, A\&A, 564, A8
\bibitem[Mart\'in-Dom\'enech et al.(2015)]{Mar15}Mart\'in-Dom\'enech, R., Manzano-Santamar\'ia, J., Mu\~noz-Caro, G.M., et al. 2015, A\&A, 584, A14
\bibitem[Mart\'in-Dom\'enech et al.(2017)]{Mar17}Mart\'in-Dom\'enech, R., Rivilla, V. M.; Jim\'enez-Serra, I., Qu\'enard, D., Testi, L., \& Mart\'in-Pintado, J. 2017, \mnras, 469, 2230
\bibitem[Mat\'e et al.(2011)]{Ma11} Mat\'e, B., Rodriguez-Lazcano, Y., G\'alvez, \'O., Tanarro, I., Escribano, R. 2011, PCCP, 13, 12268
\bibitem[Mat\'e et al.(2017)]{Ma17} Mat\'e, B., Molpeceres, G., Tim\'on, V., Tanarro, I., Escribano, R., Guillemin, J. C., Cernicharo, J., Herrero, V. J. 2017, \mnras, 470, 4222
\bibitem[Mat\'e et al.(2018)]{Mat18} Mat\'e, B., Molpeceres, G., Tanarro, I., Peláez, R. J., Guillemin, J. C, Cernicharo, J., \& Herrero, V. J. 2018, \apj, 861, 61
\bibitem[McGuire (2018)]{Mcg18} McGuire, B. A. 2018, \apjs, 219, 12
\bibitem[Meierhenrich et al.(2004)]{M04} Meierhenrich, U. J., Mu\~{no}z Caro, G. M., Hendrik B. J., Jessberger, E. K., \& Thiemann, W. H.-P. 2004, PNAS, 101, 9182
\bibitem[Meinert et al.(2016)]{M16} Meinert, C., Myrgorodska, I., de Marcellus, P. et al. 2016, Science, 352, 208
\bibitem[Mencos \& Krim(2016)]{Men16} Mencos, A., \& Krim, L. 2016, \mnras, 460, 1990
\bibitem[Mencos \& Krim(2018)]{Men18} Mencos, A., \& Krim, L. 2018, \mnras, 476, 5432
\bibitem[Milligan \& Jacox(1971)]{Mil71} Milligan, D. D., \& Jacox, M. E. 1971, \jcp, 54, 927
\bibitem[Modica \& Palumbo(2010)]{Mod10} Modica, P., \& Palumbo, M. E. 2010, \aap, 519, 22
\bibitem[Mu\~{n}oz-Caro et al.(2002)]{MC02} Mu\~{no}z Caro, G. M., Meierhenrich, U. J., Schutte, W. A., et al. 2002, \nat, 416, 403
\bibitem[Mu\~{n}oz-Caro \& Schutte(2003)]{Mu03} Mu\~{no}z Caro, G. M., \& Schutte, W. A. 2003, \aap, 412, 121
\bibitem[Mu\~{n}oz-Caro \& Dartois(2013)]{MCD13} Mu\~{no}z Caro, G. M., \& Dartois, E. 2013, Chem. Soc. Rev., 42, 2173 
\bibitem[Mu\~{n}oz-Caro et al.(2014)]{MC14} Mu\~{no}z Caro, G. M., Dartois, E., Boduch, P., Rothard, H., Domaracka, A., \& Jim\'enez-Escobar, A. 2014, \aap, 566, A93
\bibitem[Munro et al.(2012)]{M12} Munro, J. J., Harrison, S., Fujimoto, M. M., \& Tennyson, J. 2012, JPhCS, 388, 012013
\bibitem[Nomura et al.(2007)]{N07} Nomura H., Aikawa Y., Tsujimoto M., Nakagawa Y., \& Millar T. J., 2007, \apj, 661, 334
\bibitem[Nuevo et al.(2006)]{N06} Nuevo, M., Meierhenrich, U. J., Mu\~{no}z Caro, G. M., et al. 2006, \aap, 47, 741
\bibitem[Oba et al.(2016)]{Ob16}Oba, Y., Takano, Y.,  Watanabe, N., \& Kouchi, A. 2016, \apjl, 827, L8
\bibitem[\"Oberg et al.(2011)]{Ob11} \"Oberg, K. I., Boogert, A. C. A., et al. 2011, \apj, 740, 109
\bibitem[\"Oberg et al.(2015)]{Ob15} \"Oberg, K. I., Guzm\'an, V. V., Furuya, K., et al. 2015, Natur, 520, 198
\bibitem[Palumbo et al.(1998)]{P98} Palumbo, M. E., Baratta, G. A., Brucato, J. R., Castorina, A. C., Satorre, M. A., \& Strazzulla, G. 1998, \aap, 334, 247
\bibitem[Pilling et al.(2010)]{Pil10} Pilling, S., Seperuelo Duarte, E., da Silveira, E.F., et al. 2010, \aap, 509, A87
\bibitem[Pilling \& Bergantini(2015)]{PB15} Pilling, S., \& Bergantini, A. 2015, \apj, 811, 151
\bibitem[Raunier et al.(2004)]{Rau04} Raunier, S., Chiavassa, T., Marinelli, F., \& Aycard, J. P. 2004,  Chem. Phys., 302, 259
\bibitem[Ribas et al.(2005)]{R05} Ribas, I., Guinan, E.F., Gudel, M., \& Audard, M. 2005, \apj, 622, 680
\bibitem[Sandford \& Allamandola(1993)]{San93} Sandford, S. A., \& Allamandola, L. J. 1993, \apj, 417, 815
\bibitem[Sicilia et al.(2012)]{Sic12} Sicilia, D., Ioppolo, S., Vindigni, T., Baratta, G. A., \& Palumbo, M. E. 2012, \aap, 543, A155
\bibitem[Sie et al.(2019)]{NS19} Sie, N.-E., Mu\~noz Caro, G. M., Huang, Z.-H., Mart\'in-Dom\'enech, R., Fuente, A. \& Chen, Y.-J. 2019, \apj, 874, 35 
\bibitem[Solomon et al.(1971)]{Sol71}Solomon, P. M., Jefferts, K. B., Penzias, A. A., \& Wilson, R. W. 1971, \apjl, 168, L107
\bibitem[Theule et al.(2011)]{The11} Theule, P., Duvernay, F., Ilmane, A., Hasegawa, T., Morata, O., Coussan, S., Danger, G., Chiavassa, T. 2011, \aap, 530, 96
\bibitem[Unger et al.(2015)]{U15} Unger, I., Hollas, D., Seidel, R., Th\"urmer,S., Aziz, E. F., Slav\'{i}c\v{e}ck, P., 
\& Winter, B. 2015, J. Phys. Chem., 119, 10750
\bibitem[van Broekhuizen et al.(2004)]{Bro04} van Broekhuizen, F. A., Keane, J. V., \& Schutte, W. A. 2004, \aap, 415, 425
\bibitem[van Broekhuizen et al.(2005)]{Br05} van Broekhuizen, F. A., Pontoppidan, K. M., Fraser, H. J., \& van Dishoeck, E. F. 2005, \aap, 441, 249
\bibitem[Varietti \& Pimentel(1971)]{Var71} Varetti, E.L., \& Pimentel, G.C. 1971, \jcp, 55, 3813
\bibitem[Vinogradoff et al.(2013)]{V13} Vinogradoff, V., Duvernay, F., Danger, G., Theul\`e, P., Borget, F., \& Chiavassa, T. 2013, \aap, 549, A40
\bibitem[Walsh et al.(2012)]{W12} Walsh, C., Nomura, H., Millar, T. J., \& Aikawa, Y. 2012, \apj, 747, 114
\bibitem[Woon(2002)]{Wo02} Woon, D. E. 2002, \apjl, 571, L77
\bibitem[Zheng et al.(2008)]{Zhe08} Zheng, W., Jewitt, D., Osamura, Y., \& Kaiser, R. 2008, \apj, 674, 1242
\end{thebibliography}
\end{document}